\documentclass[twocolumn,prx,aps,superscriptaddress,nofootinbib]{revtex4-1}
\usepackage{latexsym}
\usepackage{palatino}
\usepackage{graphicx}
\usepackage{bm}
\usepackage{amsmath,amssymb}
\usepackage{float}
\usepackage{fnpos}
\usepackage[english]{babel}
\usepackage{array}
\usepackage[T1]{fontenc}
\usepackage[usenames,dvipsnames]{xcolor}
\usepackage{setspace}
\usepackage[compact]{titlesec}
\usepackage{amsfonts}
\usepackage{rotating}
\usepackage{threeparttable}
\usepackage{pifont}
\newcommand{\cmark}{\ding{51}}%
\newcommand{\xmark}{\ding{55}}%
\usepackage{natbib} 
\setcitestyle{super,sort&compress,comma}
\begin{document}

\title{Intramolecular Vibrational Energy Redistribution and the quantum ergodicity transition: a phase space perspective}

\author{Sourav Karmakar  and Srihari Keshavamurthy}
\affiliation{Department of Chemistry, Indian Institute of Technology, Kanpur, Uttar Pradesh 208 016, India}

\widetext

\begin{abstract}

Intramolecular vibrational energy redistribution (IVR) impacts  the dynamics of reactions in a profound way.  Theoretical and experimental studies are increasingly indicating that accounting for the finite rate of energy flow is critical for uncovering the correct reaction mechanisms and calculating accurate rates. This requires an explicit understanding of the influence and interplay of the various anharmonic (Fermi) resonances that lead to the coupling of the vibrational modes. In this regard, the local random matrix theory (LRMT) and the related Bose-statistics triangle rule (BSTR) model have emerged as a powerful and predictive quantum theories for IVR. In this Perspective we highlight the close correspondence between LRMT and the classical phase space perspective on IVR, primarily using model Hamiltonians with  three degrees of freedom. Our purpose for this is threefold. First, this clearly brings out the extent to which IVR pathways are essentially classical, and hence crucial towards attempts to control IVR. Second, given that LRMT and BSTR are designed to be applicable for large molecules, the  exquisite correspondence observed even for small molecules allows for insights into the quantum ergodicity transition. Third, we showcase the power of modern nonlinear dynamics methods in analysing high dimensional phase spaces, thereby extending the deep insights into IVR that were earlier gained for systems with effectively two degrees of freedom. We begin with a brief overview of recent examples where IVR plays an important role and conclude by mentioning  the outstanding  problems and the potential connections to issues of interest in other fields.

\end{abstract}

\maketitle

\section{Introduction}

Chemistry is all about making and breaking bonds. These fundamental processes are determined, and largely controlled, by the extent of energy flow into and out of specific modes. Consequently, the phenomenon of intramolecular vibrational energy redistribution (IVR) is the heart and soul of chemistry.
This has been recognized for over a century. For example, already in an early  work\cite{Lewis_Smith} Lewis and Smith conclude with the remark that ``At present we can only guess at the various complex factors which determine whether a molecule which has the opportunity of acquiring enough energy for activation actually does acquire it, or whether if it acquires this energy it will suffer chemical change." It is instructive, and perhaps sobering, to contrast this early remark with a more recent one\cite{Carpenteretal2016} by Carpenter, Harvey, and Orr-Ewing: ``The physical mechanisms that lead to IVR in an isolated molecule are not very different from those that lead to chemical transformation, and so, in hindsight, perhaps it was questionable whether one should have assumed that the time required for the former could be very much smaller than for the latter, as the statistical approximation requires."

One possible argument in defense of the seemingly slow progress, as evidenced by the two remarks above, is that IVR is a notoriously difficult phenomenon to come to grips with. So much so that many of the celebrated rate theories either sidestep the issue of IVR or make fairly simplistic assumptions about the nature of IVR during the reaction process. This is not entirely surprising since a mechanistic understanding of IVR is tantamount to a detailed knowledge of the reaction dynamics itself. The notoriety of IVR was further enhanced in the context of early attempts to control or manipulate reaction dynamics using lasers\cite{SchulzLee1979,crim84}. With the assumption that IVR typically happens on the timescales of a few picoseconds, there was a general consensus that performing mode-specific chemistry by controlling IVR was not a viable option. Indeed, novel suggestions for coherently controlling reactions advocated bypassing\cite{ricezhao,brumershapiro}  or beating\cite{Weidinger2009,LeeSuits2012} IVR altogether. This naturally led to suggestions for using ultra short pulses\cite{bloemzewail} and exploiting the concept of quantum interference\cite{rice01}. A different viewpoint came from optimal control theory\cite{Brif10} wherein the control fields were determined without the need for an explicit understanding of the IVR process. Nevertheless,  studies\cite{OhtsukiJCP2001,stensitzkietal18,keeferetal18} do hint towards IVR being a friend rather than a foe in the control scheme. Could it be the case that a mechanistic understanding of IVR can be a guide to uncover rational rules for designing appropriate control fields? Recent studies\cite{RafiqScholes2018,hayenkuhn19,Thomasetal19,laptevetal19} certainly indicate so. More importantly, these studies emphasize the advantage of clearly identifying the reaction coordinate and the modes that strongly couple to it. Perhaps it is fair to say that irrespective of whether one is attempting to use, bypass, or beat IVR, dynamical considerations are important. And, when the molecular dynamics  does not conform to statisticality on reaction timescales, then it is imperative to understand the dynamics of energy flow. The recent experimental\cite{LeeSuits2012} and theoretical studies\cite{ShiSchlegel2016,ShiSchlegel2019} on the field-induced fragmentation of ClCHO$^{+}$ are a good example in this context.

The relevance of IVR to control and mode-specific chemistry\cite{baer2000,DianZwier2002,DianLongarteZwier2004,Crim2008,KilleleaUtz2013,ZhaoSunGuo2015,Werdecker2018} notwithstanding, the increasing number of examples of chemical reactions exhibiting nonstatistical dynamics\cite{SunHase2002,QuijanuSingleton2011,Diau1998,baileysingleton2017,goldmancarpenter2011,glowackipilling2009,samantaschmittel2015,ussingsingleton2006,Carrascosa2017,Proenza2016,KurochiSingleton2018,ZhaoSunGuo2015,Gardner2018} has brought the focus back to characterising the IVR dynamics in molecules. Note that the term nonstatistical, in general, means that there exist dominant IVR pathways that explore only parts of the energetically available phase space due to various dynamical constraints. A classical dynamical perspective on the different dynamical constraints is provided later in sec.~\ref{CMIVR} and sec.~\ref{RRKMornot}. The literature in this regard is fairly extensive and we mention a few recent examples here. We refer the reader to several reviews\cite{rehbein_carpenter2011,Leitner2015,RehbeinWulff2015,MaHase2017,JiangGuo2019,YangHouk2018} that have appeared over the last couple of decades for a more comprehensive set of examples. An area where the importance of nonstatistical dynamics  is currently the focus of attention is in condensed phase thermal reactions of polyatomic molecules involving reactive intermediates\cite{KhoungHouk2003,OyolaSingleton2009,glowackipilling2009,QuijanuSingleton2011,goldmancarpenter2011,samantaschmittel2015,Carpenteretal2016,baileysingleton2017,KurochiSingleton2018}. Example reactions include hydrogen migration in cyclopentadiene\cite{goldmancarpenter2011}, Diels-Alder cyclizations\cite{samantaschmittel2015}, and hydroboration of alkenes\cite{baileysingleton2017}. Interestingly, the intermediates ``remember" how they were formed, resulting in product branching ratios that cannot be simply estimated based on the relative free energy barriers\cite{ThomasSingleton2008,QuijanuSingleton2011,CollinsCarpenterEzraWiggins2013,KramerWiggins2015}. These recent studies also highlight the role of energy flow into and out of the solvent degrees of freedom  and the extent to which theoretical models can a priori assume the onset of  complete thermalization due to collisions with the bath molecules. In this context,  theoretical models\cite{ZhengTruhlar2009,Glowacki2010,baileysingleton2017} have been proposed which attribute the experimentally observed branching ratios to a mix of statistical and nonstatistical effects.  Nevertheless, all such approaches need to invoke implicit assumptions about the extent of intra and intermolecular energy flow during different stages of the reaction. Although some  insights are starting to emerge into these ``semi-statistical" models, so called since they either utilize linear master equations or invoke phenomenological time scales for IVR, it is clear that ``one type may not fit all" when it comes to IVR in such complex systems, as brought out nicely by a recent study\cite{GrubbNO217} on the vibrational cooling of NO$_{2}$ in different solvents. In any case, one is still left with the fundamental issue of explaining the dynamical origins of the relative fraction of direct versus indirect trajectories\cite{rehbein_carpenter2011}.  

In most of the condensed phase examples mentioned above the issue of intra versus inter molecular IVR is obviously important. The naive expectation that solvents will ``rapidly" thermalize the intramolecular energy is just a convenient assumption and requires critical reevaluation\cite{essafiharvey18}. Indeed, as recently emphasized by Orr-Ewing\cite{OrrEwingCSR2019}, ``The chemical dynamics can appear similar to those for isolated reactive collisions in the gas-phase, even under circumstances of strong interactions of the solvent with the reacting solutes, if the solvent molecules are effectively frozen in position during a fast reaction. In this regime, the solvent molecules do not re-orient to solvate optimally the reacting species at all points along the reaction coordinate." Undoubtedly, a detailed understanding of IVR in gas phase is a prerequisite for unequivocally identifying the role of the surrounding medium in effecting intermolecular energy flow and hence a mechanistic understanding of most of the chemical reactions. As an example we mention efforts  to dissect the intra versus intermolecular energy flow pathways that have led to deep insights into the hydrogen bond dynamics in bulk water as well as at the air-water interface\cite{HsiehBonn2013,PostBonn2015,JeonBonn2017,demarcoetal13,kananenkaetal18,ramaseshaetal13}.

This brings us to the main issue that is of interest to us - do we fully understand IVR dynamics in isolated molecules? From the brief set of examples mentioned below it will hopefully become clear that although  considerable advances have been made towards characterizing the IVR pathways in isolated molecules, claiming that we fully understand gas phase IVR dynamics would be premature. For instance, we still do not have an answer to even the seemingly simple question regarding the necessary and sufficient conditions that would guarantee a gas phase unimolecular reaction to be in the Rice-Ramsperger-Kassel-Marcus (RRKM) regime i.e., a regime wherein the IVR timescale is sufficiently short that dynamical considerations can be bypassed and statistical arguments can be invoked for calculating rates (cf. sec.~\ref{RRKMornot} for details). Interestingly, in this context, there are connections to the recent surge of activity in the condensed matter physics community on the topics of thermalization\cite{rigol2008,Eisert2015,alessio2016} and many-body localization (MBL)\cite{NandHuse2015,Altman2018}. The eigenstate thermalization hypothesis (ETH)\cite{Deutsch1991,Srednicki1994,Tasaki1998,alessio2016} and MBL continue to pose fundamental questions that relate to the notions of ergodicity (classical and quantum), entanglement, and persistence of quantum coherence. Exploring the connections between these different fields seems to be a tantalizing opportunity for fresh perspectives on the IVR phenomenon.

Gas phase systems have always been a favorite playground for physical chemists to search for mode-specific effects and nonstatistical dynamics\cite{SunHase2002,Marcus2013,lourderaj2009,Smith04,Beck03,Leitner2003,Dian08,Pate2008}. Indeed, studies spanning several decades have established that a mechanistic understanding of IVR is essential for interpreting reaction dynamics\cite{SunHase2002,QuijanuSingleton2011} as well as explaining the intricate features in the overtone spectra of molecules\cite{BaggotMills1986,JacobsonField1999,Callegari2000,Ishiuchi2006}. The number of examples which exhibit non-RRKM behaviour continues to increase, leading ever so slowly towards the holy grail of associating specific structural motifs with possible universal classes of IVR dynamics. For instance, low barrier isomerization reactions\cite{baer2000,DianZwier2002,DianZwier2004,Zwier2006,Dian08,StevensTroe2009,Leitner2003} are now expected to be non-RRKM. Recently, the ground breaking experimental technique of dynamic rotational spectroscopy\cite{Dian08}  by Pate and coworkers has provided an unique opportunity to study the rich dynamics associated with the isomerization reactions. On the other hand, presence of one or more centers of flexibility leading to large amplitude motions can result in significant enhancement of IVR. For example, substantial progress has been made recently towards uncovering the role of methyl rotors\cite{Gascooke2015,Gardner2018,Tuttle2019} in enhancing IVR with the vibration-torsion coupling being the key parameter. Nevertheless, the issue has its share of subtleties. In particular, molecules with planar conformations undergo slower IVR from acetylinic C-H stretch fundamentals in comparison to those with non-planar conformations\cite{engelhardtetal2001}. In fact, considerable amount of work\cite{McIlroyScoles1994,kerstel1994,andrewsetal98,hudspethetal98,yoopate1,yoopate2,YooPate2004} has been done to characterize the IVR from acetylinic C-H stretches in molecules with and without large amplitude modes. As a further example, extensive studies\cite{BoyarkinPerry1999,twagirayezuetal10,sibertjcp19} of IVR in methanol illustrates the complexity of energy flow, even in this relatively small molecule, with the existence of several time scales ranging from a few hundred femtoseconds to several picoseconds. 

From a dynamical standpoint, the nature and extent of IVR is linked to the avoidance of deep minima on potential energy surfaces\cite{lourderaj2009,manihase12}, and hence the minimum energy path\cite{Maedaetal2015,tsutsumi18}, birth of new vibrational modes\cite{ishikawaHCP1999,Kellman2007,Farantos2009}, and existence of approximately conserved quantities called as polyads\cite{krasnostepanov13} (cf. sec.~\ref{CVPTpolyads} for a brief explanation). Thus, gas phase nucleophilic S$_{\rm N}2$ reactions can exhibit post-transition state non-RRKM dynamics\cite{SunHase2002} (see also the work of Craig and Brauman\cite{CraigBrauman}) with the emergence of new mechanistic pathways as a function of collision energies\cite{SzaboCzako2015}.  Knowledge of IVR timescales is also crucial towards generalizing Polanyi's rules\cite{Polanyi1972}, regarding the relative efficiency of translations versus vibrations in promoting a reaction, in terms of the sudden vector projection model proposed by Guo and coworkers\cite{JiangGuo2013,Stei2018}. Interestingly, recent works\cite{zhaomantheguo,ellerbrockmanthe} show that one also requires information on the IVR pathways in order to rationalize the reaction probabilities in initial state-selected  collisions involving polyatomic species. At the same time it is becoming increasingly clear that information on the energy flow pathways  at various levels of detail is essential in order to explain the origin of mode and bond specific effects in gas-surface reactions\cite{KilleleaUtz2013,Werdecker2018,JiangGuo2014,JiangGuo2016,JiangGuo2019,shiretal18}. A further impetus for a detailed understanding of IVR in isolated molecules has to do with the interest in characterizing the flow of heat through molecules that act as junctions in nanoscale devices\cite{Segal2016,PandeyLeitner2016}. In this context, experiments do show that structural changes can lead to unidirectional IVR, mimicking a vibrational energy diode, and violations of Fourier heat law\cite{peinetal13,rubtsovburin19}.

The examples above are a testimony to the richness of the phenomenon of IVR, and at the same time a reminder of the complexity of it as well. The inevitable question then is: are there are any overarching universal features of energy flow dynamics that one can extract from the innumerable studies or one has to treat every system in its own right? If all the molecules conformed to the RRKM assumption of random and fast IVR then that would certainly be a comforting, albeit a rather dull, level of universality. This would also negate any possibility of active control. At the other extreme, if every molecule had its own unique IVR dynamics then all hopes of universality are gone. Nevertheless, activation of specific modes could then be exploited in many interesting ways. Surely, the examples cited above indicate that there is hope of extracting certain universal features despite the nonstatistical nature of the IVR dynamics. The key to discovering these universal features lies in the couplings that mediate energy flow between two or more modes in the molecule. Partially, the reason for such an expectation is that  molecules, by definition, have a local coupling structure. Thus, one might anticipate the transferability of the local couplings involving a specific functional group or structural motif from one molecule to another similar molecule. Combined with the  fact that the couplings responsible for IVR are expected to obey certain scaling relations, this ensures a universal IVR dynamics over relatively short time scales. Of course, at longer times the difference in the nature of the coupling beyond the immediate neighbours will distinguish one molecule from the other. Indeed, exploiting this feature is central to the relaxation-assisted 2D infrared spectroscopy technique\cite{Kurochkin2007}. However, if active control is what one is after then the short time universal IVR dynamics is hopefully the relevant part. Alternatively, if the field-free IVR pathways involving modes that couple to the reaction coordinate are clearly identified, one could attempt to make rational  changes to the molecular structure in order to bias a specific pathway over the rest. Examples in this direction include the recent studies by Rafiq et al\cite{RafiqScholes2018}, Schmitz et al\cite{Schmitz2019}, and Delor et al\cite{DelorScience2014,DelorNatChem2015,DelorNatChem2017} wherein modulating specific anharmonic resonances  can result in exquisite control.   

Although there are many sources of mode-mode couplings that lead to IVR, the ones  that are particularly to the IVR process are the anharmonic resonances. Essentially, an anharmonic resonance signals significant interactions that arise due to two or more modes having frequencies that are nearly commensurate with each other. Examples like the Darling-Dennison resonance and Fermi resonance have played an important role in understanding the spectral features of several molecules. A famous example of Fermi resonance is the carbon dioxide molecule where the first overtone $2\nu_{2}$ of the degenerate bending mode ($\nu_{2} \sim 667$ cm$^{-1}$) is nearly commensurate with the fundamental of the symmetric stretching vibration ($\nu_{1} \sim 1337$ cm$^{-1}$). Similarly, assigning the spectrum of the water molecule requires taking into account both the $1:2$ Fermi resonances between the stretching and bending modes and the $2:2$ Darling-Dennison resonances between the stretches. More generally, resonances can involve more than two modes as in, for example,  acetylene\cite{Kellman1990,RoseKellman1995,TyngKellman2006} and thiophosgene\cite{StricklerGrub2004,JungTaylorSibert2006,sibertgrub2006}. Typically, depending on the energy at which an initial nonstationary or zeroth-order bright state (ZOBS) is experimentally prepared, several of these anharmonic resonances can participate in the IVR dynamics. Spectroscopically, the signatures are found in terms of complicated intensity patterns (``intensity borrowing") due to the fractionation of specific transitions. What is, however, important to note is that despite the spectral complexity it is rare for the spectra to be completely random. In particular, several careful studies have established that there are spectral patterns that can be recognized due to the existence of approximate conserved quantities called as polyads. Indeed, as described in detail in the recent perspective by Hermann and Perry\cite{HermanPerry2013}, effective Hamiltonians based on such polyads have been of immense value in understanding the overtone spectra of numerous molecules. 

\subsection{Motivating this perspective}

The existence of polyads implies dynamical bottlenecks to thermalization. Where do these bottlenecks come from?  A second set of questions arises from the issue of what one means by ``modes" in highly excited molecules.  Although the ZOBS can be intuitively described in terms of some appropriate zeroth-order vibrational modes, it is not necessary that the subsequent dynamics is easily understood in terms of the same zeroth-order modes. In fact, the other consequence of anharmonic resonances is that entirely new set of motions can emerge with varying energies and polyads\cite{TyngKellman2006,Kellman2007}. The normal-to-local transition\cite{JaffeBrumer1980,millsrobiette85} in triatomic molecules is one such instance. The appearance of local-bender and counter-rotating motion in acetylene bending dynamics is another example\cite{FieldAcetylene2000} which has been experimentally studied in great detail. Similarly, the manifestation of isomerization modes in HCP and other such small molecules has been confirmed in terms of the spectral perturbations\cite{BeckHCP1997,ishikawaHCP1999}. A key point to note here is that this appearance and disappearance of modes, and the existence of polyads, is a hallmark of nonlinear classical dynamics. 

The observations above, therefore, raise an interesting question.  To what extent can the time and frequency domain IVR results  be understood from a purely classical dynamical perspective?  The question of the extent of ``quantumness" of IVR is worth pondering over in light of  recent studies that highlight the subtleties involved in identifying phenomena that are of purely quantum origin. For instance, the so called quantum speed limit, that puts bounds on the rate at which an initial quantum state can evolve, survives the classical ``$\hbar \rightarrow 0$" limit\cite{delcampo2018,Ohzeki2018}. Another example comes from the work\cite{FrancoBrumerPRL2006} of Franco and Brumer wherein it was established that the notion of quantum coherent control, traditionally ascribed to quantum interference induced by the relative phases between two or more ac-fields, also survives the classical limit. Undoubtedly, quantum effects such as tunneling, zero-point energies, and entanglement are expected to play a role in certain regimes and obviously important to account for. However, as stated by Heller in the preface to his recent book\cite{hellerbook}, ``Even the smallest coherent quantum systems yield their best secrets aided by an advanced knowledge of classical mechanics and semiclassical connections". Along similar lines, in a recent note\cite{millerjcp2012}, Miller beautifully summarizes the issue as ``The point of this essay, therefore, is that coherence effects may be of quantum or classical origin and that it is not always so obvious which it is...To make matters even more ambiguous, sometimes whether the observed coherence is quantum or classical depends on what is being observed!". Indeed, we now have studies on systems ranging from molecules to trapped Bose-Einstein condensates that have established that even phenomena such as dynamical tunneling\cite{hellerdavis81,Hensinger2001,Steck2001,Chaudhury2009,Oberthaler2005,dyntunbook,tomsovicbook,Ziboldetal10,Bohigas1993}, entanglement\cite{FuruyaPRL1998,ArulPRE2001,Chaudhury2009,Neill2016}, and quantum coherence\cite{Zurek2001} are influenced by structures in the classical phase space. Consequently, independent of the extent of quantitative agreement, a detailed understanding of the classical dynamics is crucial for establishing a clear and unambiguous baseline for identifying  genuinely quantum effects.

Thus, given the prevailing and highly useful view of intuitively thinking about molecules in terms of the ``ball-and-spring" models,  it should not come as a surprise that the field of IVR has benefited immensely from classical-quantum correspondence studies. In this context, the literature is fairly vast, and several pioneering works\cite{uzer91,ezra1998,davis1995,toda2005,TodaACP2002,Kellman2007,Farantos2009} have amply illustrated the role of nonlinear (anharmonic) resonances in classical (quantum) IVR dynamics. Detailed insights into the nature and mechanism of IVR have emerged from dynamical concepts such as partial barriers in phase space formed by cantori\cite{Kadanoff1984,Geisel1986,BrownWyatt1986,MackayMeiss1988,MaitraHeller2000,matthiasetal12} and vague tori\cite{ShirtsReinhardt1982,Haseetal1984}, overlapping of resonances generating chaos\cite{Chirikov79,OxtobyRice,ramaswamymarcus81}, and bifurcations of periodic orbits leading to the birth of new modes\cite{ishikawaHCP1999,Farantos2009,TyngKellman2006,Kellman2007,TyngKellman2010}. The corresponding signatures are certainly present in the complex structure of the eigenstates, and manifest in the frequency domain high resolution spectra\cite{lehmannetal94,HermanPerry2013,manikandan2009,keskepate00}. Moreover, given that IVR is relevant to determining the correct reaction mechanisms and rates, it is now incontrovertible  that transition state theory (TST)\cite{eyring1935,wigner1938,WaalkensTST2007} and the statistical RRKM theory\cite{baerhasebook,riceramsperger1927,kassel1927,marcus1952} are best understood in terms of transport in the classical phase space. For instance, the central objective in TST of constructing a recrossing-free dividing surface is ideally achieved from a phase space perspective\cite{wigner1938,Pechukas1973,PollackPechukas1978,WaalkensTST2007}. The recent progress on understanding the roaming mechanism and rates from a phase space perspective\cite{WiggingRoaming2017} is a testimony to the power of such an approach. Similarly, the crucial assumption of ``instantaneous" IVR in the RRKM theory is related to the concept of ergodicity on the constant energy surface in phase space. Furthermore, since chaotic trajectories can be non-ergodic, theories that rely on modifying RRKM in terms of the classical phase space being ``divisible" into regular and irregular parts need to invoke subtle assumptions concerning transport in the multidimensional phase space and the nature of the post-TS IVR dynamics - assumptions which have rarely been subjected to the necessary level of scrutiny. 

So, how detailed an understanding of the classical phase space transport is required to gain insights and, hopefully, exploit the IVR process? The short answer is that information from classical trajectories at all levels of detail is useful. But, how does one extract the relevant dynamical information? Much of the deep insights into IVR and reaction dynamics from a classical-quantum correspondence viewpoint has come from systems with effectively two degrees of freedom\cite{uzer91,ezra1998,reinhardt82,Farantos2009,davis1995,toda2005,sibert1jcp82}. Since even a triatomic molecule has three vibrational degrees of freedom, a particularly pressing issue has to do with extending our understanding of classical phase space transport, and the implications for the corresponding quantum dynamics, in systems with at least three degrees of freedom.  IVR is the mechanism by which energy is funneled into the reaction coordinate and hence access to the transition state. At the same time, post-TS IVR dynamics largely decides the selectivity and branching ratios of reactions\cite{SunHase2002,LourderajHase2008,ThomasSingleton2008,SiebertTantilloHase2011,CollinsCarpenterEzraWiggins2013,KramerWiggins2015}. Since the classical phase space structures associated with a TS are necessarily connected with structures that are responsible for IVR, both pre and post TS, one cannot hope to uncover the true reaction mechanism without a clear understanding of the  global transport in the multidimensional phase space. However, there are several technical as well as conceptual challenges. Nevertheless, over the past decade, considerable progress has been made in the context of TST, in terms of constructing optimal locally recrossing free dividing surfaces for systems with three or more degrees of freedom\cite{wiggetalprl01,Uzeretal2002,WaalkBurbWigJCP2004,WaalkensWiggins2004,LiTodaTamikiJCP2005,jaffeetal05}.  These advances have led to crucial insights into TST and have made it eminently clear that the adopting a phase space perspective is no longer optional as one goes beyond two degree of freedom systems. Given that these beautiful studies have provided the higher dimensional generalization of TST, similar levels of detailed insights into the pre and post-IVR dynamics are now needed to obtain an accurate description of the reaction mechanism. This is the focus of this perspective.

We begin with a brief description of the IVR process and summarize the various models. In particular the so called state space model of IVR will be of particular interest. This is then followed, after a brief motivation, by the classical dynamical setting for IVR in systems with several degrees of freedom. We then highlight recent advances in our understanding of the influence of multiple anharmonic (nonlinear) resonances to the IVR process and the consequences for gas phase unimolecular reactions. A key point that we highlight is that, counter to intuitions, one can have stabilization even in the presence of several anharmonic resonances. We end with a brief summary and discussion of the key outstanding issues. 

\section{Models for IVR}
\label{IVRmodels}

\subsection{Choice of Hamiltonian}
\label{IVRHams}
In principle, one should start with the full Wilson-Howard Hamiltonian\cite{WilsonHoward1936} for a molecule which would include the entire gamut of couplings that influence the IVR process.  There are certainly examples\cite{GambogiScoles1994,McIlroyScoles1994,kryvohuzmarcus1,kryvohuzmarcus2} where rotational-vibrational  and coriolis couplings are essential for a complete understanding of the energy flow dynamics. Here, however, we will work with simpler Hamiltonians that only account for the vibrational couplings. 

Consider a $N$-atom polyatomic molecule with $f \equiv 3N-6$ vibrational degrees of freedom. A possible Hamiltonian describing the vibrational motions can be expressed as an expansion
\begin{equation}
    H({\bf P},{\bf Q}) = \sum_{j=1}^{f} \frac{1}{2} \left(P_{j}^{2} + Q_{j}^{2} \right) + \sum_{j,k,l=1}^{f} F_{jkl} Q_{j} Q_{k} Q_{l} + \ldots
\end{equation}
with $({\bf P},{\bf Q})$ being the dimensionless normal mode momenta and coordinates. In the above we have assumed that the equilibrium corresponds to ${\bf Q} = {\bf 0}$, with the terms $Q_{j}Q_{k}Q_{l}$ and higher representing the mode anharmonicities and coupling between the various modes. The coupling strengths $F_{jkl}$, for instance, are related to the derivative $(\partial^{3}V({\bf Q})/\partial Q_{j} \partial Q_{k} \partial Q_{l})_{\bf 0}$ of the multidimensional potential energy surface $V({\bf Q})$. Note that there are several other ways to represent the Hamiltonian and, in particular, one can adopt a local mode perspective wherein the individual vibrations are taken to be anharmonic. On the other hand, in internal bond coordinates the mode-mode couplings appear in the momentum space as well. The various representations are related\cite{KellmanJCP1986} and for a given molecule a specific representation may be more suitable than others. We do not go into the details here, but refer the reader to several excellent articles\cite{JaffeBrumer1980,KellmanJCP1986} that bring out the pros and cons of the various representations. In any case, for the purpose of this article, we consider Hamiltonians of the form
\begin{equation}\label{gen_ham}
     H({\bf P},{\bf Q}) = H_{0}({\bf P},{\bf Q}) + \sum_{j \neq k \neq l}^{f} F_{jkl} Q_{j} Q_{k} Q_{l} + \dots
\end{equation}
with $H_{0}$ describing the uncoupled motion of a set of anharmonic oscillators. Such Hamiltonians are expected to be suitable for describing the dynamics at high vibrational excitations. 

A useful way to understand the influence of the coupling terms is to use the transformation 
\begin{equation}
    {\bf Q} = \frac{1}{\sqrt{2}} ({\bf a} + {\bf a}^{\dagger}) \,\,\,\, ; \,\,\,\, {\bf P} = \frac{i}{\sqrt{2}} ({\bf a}^{\dagger} - {\bf a})
\end{equation}
involving the creation (${\bf a}^{\dagger}$) and destruction (${\bf a}$) operators. The Hamiltonian can now be expressed as
\begin{equation}
    H = \sum_{j=1}^{f} {\cal E}_{j}(v_{j}) + \sum_{j \neq k \neq l}^{f} \Phi_{jkl} (a_{j}^{\dagger} + a_{j}) (a_{k}^{\dagger} + a_{k}) (a_{l}^{\dagger} + a_{l}) + \ldots
\end{equation}
with $v_{j} \equiv a_{j}^{\dagger} a_{j}$ being the number operator which yields the number of excitation quanta in the $j^{\rm th}$-mode. The anharmonic energy associated with the $j^{\rm th}$-mode is denoted by ${\cal E}_{j}$ and the couplings $\Phi_{jkl}$ are related to the cubic couplings $F_{jkl}$. Note that the representation above is also ideal for performing canonical Van-Vleck perturbation theory (CVPT)\cite{VanVleck1951,Sibert1990,JoyeuxSugny2002} - a key approach for constructing effective Hamiltonians and extracting information about dominant resonances and polyad numbers. The eigenstates $|{\bf v}\rangle$ of the zeroth-order Hamiltonian are conveniently labeled by the quantum numbers $|v_{1},v_{2},\ldots,v_{f}\rangle$ with the corresponding eigenenergies, and thus can be thought of as points in the discrete $f$-dimensional quantum number space (QNS). The various mode-mode coupling terms connect (mix) a given $|{\bf v}\rangle$ with other zeroth-order states $|{\bf v}'\rangle$, lying a distance ${\cal Q} \equiv \sum_{j=1}^{f} |v_{j}-v_{j}'|$ away in the QNS. For example, the cubic terms connect $|{\bf v}\rangle$ with all other states located at a distance ${\cal Q}=3$. Before going on to describe the models for IVR, it is useful to point out that the Hamiltonian can be written in a more compact notation as
\begin{equation}
    H = \sum_{j=1}^{f} {\cal E}_{j}(v_{j}) + \sum_{{\bf m}} \Phi_{\bf m} \prod_{j=1}^{f} \left(a_{j}^{\dagger} + a_{j}\right)^{m_{j}}
    \label{hamcompact}
\end{equation}
with ${\bf m} = (m_1,m_2,\ldots,m_f)$ and $\sum_{j} m_{j}$ being the order of the various coupling terms. Thus, for instance, $\sum_{j} m_{j} = 3$ represent all the cubic terms. Another useful way of expressing the Hamiltonian is as follows
\begin{equation}
    H = \sum_{\bf v} \epsilon_{\bf v} |{\bf v}\rangle \langle {\bf v}| + \sum_{{\bf v} \neq {\bf v}'} V_{{\bf v} {\bf v}'} |{\bf v}\rangle \langle {\bf v}'|
\end{equation}
with the matrix element $ V_{{\bf v} {\bf v}'}$ being related to the coupling coefficients $\Phi_{\bf m}$.
The form above is closely connected to model Hamiltonians that have been investigated extensively by the condensed matter physics community. In particular, given that $\epsilon_{\bf v}$ and $V_{{\bf v} {\bf v}'}$ can be interpreted as ``site" energy and ``hopping" terms, the similarity to the Bose-Hubbard model and the tight binding models means that the phenomenon of IVR has connections to the phenomenon of Anderson localization\cite{Anderson1958}, thermalization\cite{rigol2008,Eisert2015,alessio2016}, and many-body localization\cite{NandHuse2015,Altman2018}. Indeed, these analogies have played a crucial role in formulating the local random matrix theory (LRMT)\cite{schofieldwolynes1994,leitnerwolynes1997} model for IVR. It is worthwhile mentioning that IVR is an example of transport in an interacting many body system.

\subsection{Analyzing the IVR dynamics}
\label{IVRmeasures}

Given the form of the Hamiltonian, preparing an initial nonstationary state of interest corresponds to creating a energy hot spot in the molecule. The manner in which this initial hot spot redistributes throughout the molecule due to the various mode-mode couplings i.e., the subsequent dynamics of the initial state is the phenomenon of IVR. For a pedagogical introduction and an excellent overview of the IVR process we refer the reader to the comprehensive review by Nesbitt and Field\cite{nesbittfield96}. Here we provide a rather short summary of the various quantities of interest in IVR studies. Consider a ZOBS $|{\bf v}\rangle$ that has been prepared initially.  
In the presence of the cubic and higher order couplings the nonstationary ZOBS will  interact (mix) with other dark zeroth-order states resulting in IVR. 
The IVR dynamics of the nonstationary state $|{\bf v}\rangle$ can be probed via several different quantities. From a time-dependent viewpoint, the survival probability
\begin{equation}
    P_{{\bf v}}(t) \equiv |\langle {\bf v}(0)|{\bf v}(t) \rangle|^{2} = |\langle {\bf v}(0)|e^{-iHt/\hbar}|{\bf v}(0) \rangle|^{2}
\end{equation}
is a measure of the extent of IVR out of the state. A time-independent measure of the extent of IVR, resulting in the mixing of the various zeroth-order states, is the dilution factor\cite{macdonald83,sibertgrub2006}
\begin{equation} \label{df_def}
    \sigma_{\bf v} = \sum_{\alpha}|\langle \alpha|{\bf v} \rangle|^{4} \equiv N_{\rm eff}^{-1}(|{\bf v}\rangle)
\end{equation}
where $|\alpha \rangle$ denotes the eigenstates of the full Hamiltonian.
Note that $\sigma_{\bf v}$ is, assuming that the off-diagonal terms can be neglected, the infinite time limit of $P_{\bf v}(t)$.
The dilution factor indicates the extent of fragmentation of the initial state due to the various mode-mode couplings. Basically, $N_{\rm eff}$ measures the number of eigenstates that have the ZOBS character. In other words, this is the number of states over which the intensity of the initial ZOBS is shared or borrowed. Consequently, the full complexity of the IVR process is encoded in the eigenstates of the full Hamiltonian.  In principle, therefore, once the survival probability and other cross-probabilities have been determined, it is possible to decipher the IVR occurring in the system. Complementary to the time-dependent viewpoint, insights into the nature of the eigenstates at specific energies of interest, in terms of the energy splittings and relative intensities measurable via high-resolution frequency domain spectroscopy, can yield insights into the IVR pathways. 

\subsubsection{Computational challenges and perturbative viewpoints}
\label{CVPTpolyads}

For molecules with several degrees of freedom computing the survival probability or ``deconstructing"  the molecular eigenstates  are daunting tasks.  Gaining insights into a specific set of eigenstates, given the visualization constraints associated with multidimensional wavefunctions and the fact that they encode the long time IVR, is relatively less easy.
Nevertheless, assuming the availability of sufficiently accurate multidimensional potential energy surfaces\cite{puzzarinietal18},  powerful approaches do exist for computing the eigenstates and survival probabilities in an efficient manner for large molecules. As examples we mention the multi-configuration time-dependent Hartree (MCTDH) method\cite{BeckMCTDH2000,IungMCTDH2004,GattiMCTDH2004,PasinMCTDH2006,MeyerMCTDH2006,PasinMCTDH2008}, the recursive residue generation method (RRGM)\cite{WyattRRGM1983,WyattRRGMACP,MinehardtWyatt1999}, the permutationally invariant polynomials (PIP) approach\cite{BowmanPIP2009,BowmanPIP2011,BowmanPIP2018}, method of filter diagonalization\cite{Neuhauser1990,Neuhauser1995,MandelshtamPRL1997}, iterative solvers using various preconditioning techniques\cite{thomasetaljcp18,carrington18}, and methods based on phase space truncation schemes\cite{PoirierJCTC2003,PoirierIIJCP2004,PoirierIIIJCP2004,PoirierACP}. A cautionary note, however, needs to be made. Fitting  the different levels of ab initio data to obtain the multidimensional potential energy surfaces can result in different dynamics. Sometimes, as recently shown by Perry et al\cite{PerryThompson2013} in their detailed study of the dissociation dynamics of HO$_{2}$, the extent of regular and chaotic dynamics can differ considerably.  
An altogether different perspective, one that avoids the need for a prior knowledge of the potential energy surfaces, is based on the so called ``on the fly" ab initio molecular dynamics approach. Although ab initio MD is a powerful approach, for moderately sized gas phase molecules one is still limited to a relatively small number of trajectories. Moreover, assuming the phase space to be mixed regular-chaotic, it is questionable whether even a few thousand trajectories are sufficient to sample the constant energy surface in the phase space with several tens of degrees of freedom. 

Yet another powerful approach is based on using CVPT to perturbatively transform the Hamiltonian to an effective Hamiltonian $H \equiv H_{0} + V$ with
\begin{equation}
    H_{0} = \sum_{j=1}^{f} \omega_{j} \left(v_{j} + \frac{1}{2} \right) + \sum_{j \leq k}^{f} x_{jk} \left(v_{j} + \frac{1}{2} \right)\left(v_{k} + \frac{1}{2} \right) + \ldots
    \label{dunham}
\end{equation}
being the Dunham Hamiltonian and $V$ containing the anharmonic resonances that need to be accounted for in order to accurately determine the eigenstate energies. In the above, $\omega_{j}$ is the harmonic frequency associated with the $j^{\rm th}$-mode and $x_{jk}$ represent the anharmonicities. Consider an example model $3$-mode system wherein the CVPT analysis may uncover that  three anharmonic resonances need to be included. Assume, without reference to any specific molecule, that the resonances are
\begin{eqnarray} \label{modelres}
 V_{1} &=& (a_{1}^{\dagger})^{2}a_{2} + a_{1}^{2} a_{2}^{\dagger} \nonumber \\
 V_{2} &=& (a_{1}^{\dagger})^{3} a_{2}^{2} + a_{1}^{3} (a_{2}^{\dagger})^{2} \\
 V_{3} &=& a_{2}^{\dagger} a_{3}^{2} + a_{2} (a_{3}^{\dagger})^{2} \nonumber
\end{eqnarray}
For the rest of this article we will use the model Hamiltonian proposed by Martens\cite{ccmjstat1992}
\begin{equation}
    H_{\rm model} = 
 \sum_{i=1}^{3} \bigg[\omega_i \Big({a}_i^{\dagger}  {a}_i + \frac{1}{2}\Big) + \frac{1}{2} \alpha_i \Big({a}_i^{\dagger} {a}_i + \frac{1}{2}\Big)^2 \bigg] + \sum_{i=1}^{3} \beta_{i} V_{i}
\label{ham_qm_ccm}
\end{equation}
with the resonances as in eq.~\ref{modelres} to illustrate various levels of classical-quantum correspondence in the IVR dynamics. Note that in $H_{\rm model}$ we have only included diagonal quadratic anharmonicities and is therefore a special case of the more general Dunham expansion given above. Nevertheless, this is not a severe limitation since the higher order anharmonicities are much smaller in comparison and for IVR the various anharmonic resonances $V_{j}$ are the key terms. The parameters of the zeroth-order Hamiltonian are given in Table~\ref{tbl:example}. In addition, for convenience, the notation $(\beta_1,\beta_2,\beta_3) \times 10^{-5} \equiv [\beta_1,\beta_2,\beta_3]$ will be used throughout this work.

Note that if only $V_{1}$ is necessary then $v_{3}$ is a good quantum number (conserved), implying that mode-$3$ is not involved in IVR. However, $v_{1}$ and $v_{2}$ are no longer good quantum numbers and energy flow can occur between the two modes. Nevertheless, the quantity $P \equiv v_{1} + 2 v_{2}$ is still a good quantum number. This is an example of a polyad, and in such a case the Hamiltonian has effectively $f=1$ and is block-diagonal in $P$. Similarly, if both $V_{1}$ and $V_{3}$ are present then neither the zeroth-order quantum numbers $(v_{1},v_{2},v_{3})$ nor the polyad $P$ are conserved and IVR can involve all the modes. Even in this case one has the exact conservation of the ``super" polyad $P' \equiv v_{1} + 2v_{2} + v_{3}$ and the Hamiltonian is block-diagonal in $P'$ with $f=2$ effectively.  On the other hand, inclusion of the $V_{2}$ term along with $V_{1}$ and $V_{3}$ leads to the breakdown of the constancy of $P'$ as well. In other words, $V_{2}$ acts as a polyad-breaking term in the model Hamiltonian. Now, one has to deal with the full three degrees of freedom aspect of the Hamiltonian. The advantages of such effective Hamiltonians are manifold  and have proved rather useful in the context of understanding the complex spectral patterns in a number of molecules. More importantly, the CVPT analysis identifies the key resonances that give rise to the various IVR pathways in the system\footnote[2]{It is also worth emphasizing at this stage that the CVPT procedure has close correspondence to the normal form perturbation theory in classical mechanics. In fact the effective Hamiltonians can be considered as resonant normal forms which are an important starting point for studying stability and transport in dynamical systems theory. For an excellent introduction, refer to the paper by Fried and Ezra\cite{FriedEzra1987} and Waalkens, Schubert, and Wiggins\cite{WaalkensTST2007}.}. A fairly exhaustive account of the various effective Hamiltonians can be found in the perspective by Herman and Perry\cite{HermanPerry2013}.

Undoubtedly, as further progress in theoretical methodologies open up the possibility of investigating the IVR dynamics in increasingly larger molecules, rationalizing the computational results would still be a difficult enterprise.  
What one requires here are measures that capture the locality of IVR as well as models which can aid in uncovering the universal aspects of the IVR dynamics. In addition, exploiting the recent advances in our understanding of classical dynamics, along with modern techniques from nonlinear dynamics, would be required for systems with several degrees of freedom.

\subsubsection{``Tiering" the IVR process}

\begin{figure}
\centering
  \includegraphics[width=1\linewidth]{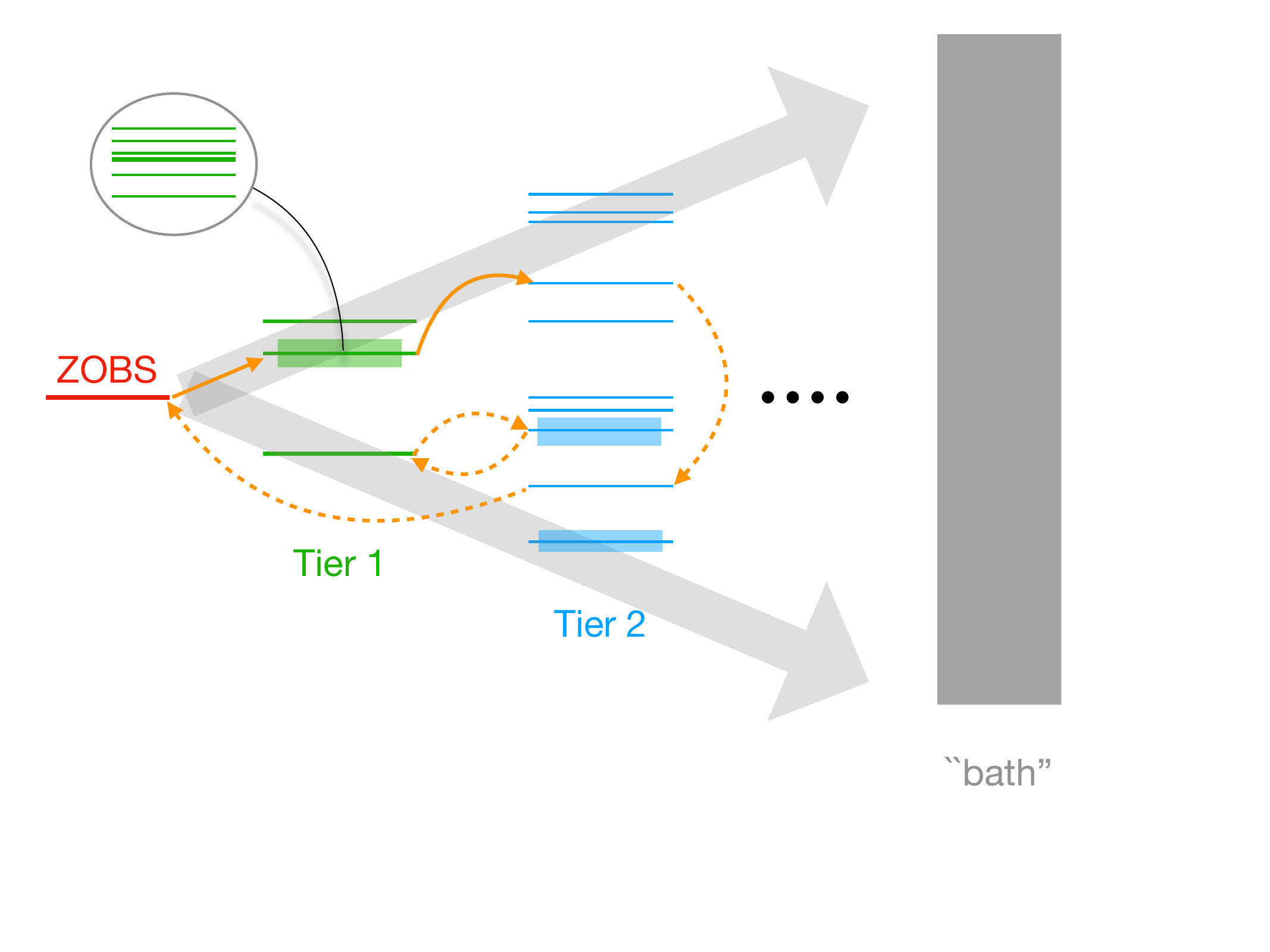}
  \caption{A schematic of the tier model. The dashed orange lines indicate possible backflows resulting in ``loops" in the tier picture. The ``fuzzy" levels (highlighted by grey circle) represent a highly degenerate set of states that could get involved provided Arnold diffusion is relevant. See text for details. }
  \label{tiersketch}
\end{figure}

Although $N_{\rm eff}$ has been widely used in many studies of gas phase IVR, a more crucial parameter has to do with the local density of states that couple to the ZOBS. A convenient measure, among other possible choices, for this is the quantity\cite{mgrub99,mgrub00}
\begin{equation}
    N_{\rm loc}(|{\bf v}\rangle) = \sum_{{\bf v}'} {\cal L}_{{\bf v}{\bf v}'}^{2}
\end{equation}
where we have denoted
\begin{equation}
    {\cal L}_{{\bf v}{\bf v}'} =  \frac{1}{\sqrt{1 + \left(\frac{(E^{0}_{\bf v} - E^{0}_{{\bf v}'})}{\langle {\bf v}|V| {\bf v}' \rangle}\right)^{2}}}
    \label{pertind}
\end{equation}
with $E^{0}_{\bf v} \equiv \sum_{j=1}^{f} {\cal E}_{j}(v_j)$. The quantity ${\cal L}_{{\bf v}{\bf v}'}$ is a perturbative indicator of the extent of coupling of a given ZOBS $|{\bf v}\rangle$ with the other zeroth-order states. In fact, ${\cal L}_{{\bf v}{\bf v}'}$ can be used to interpret the IVR in terms of the tier-model\cite{Freed1976,MuthukumarRice1978,stannardetal81}. In this model, as shown in a qualitative sketch in Fig.~\ref{tiersketch}, zeroth-order states $|{\bf v}'\rangle$ that are optically dark but with ${\cal L}_{{\bf v}{\bf v}'}$ values above a preset threshold form the set ${\cal T}_{1}$ of first tier ``doorway" states that couple to the ZOBS $|{\bf v}\rangle$. Note that the set of states in ${\cal T}_{1}$ need not all be at the same distance ${\cal Q}$ in the QNS. Next, the zeroth-order states that are significantly coupled, again based on eq.~\ref{pertind}, to the states in ${\cal T}_{1}$ form the second tier ${\cal T}_{2}$. Repeating the procedure unfolds the states involved in the IVR dynamics into a hierarchical set of tiers, with the density of states increasing with the increasing tier level. One can also use different tiering criteria to sort out the tiers. For example, instead of using ${\cal L}_{{\bf v} {\bf v}'}$ the tiers can be organized by the QNS distance ${\cal Q}$. In any case, the tier picture of IVR emphasizes the importance of the local density of states as opposed to the total density of states at the energy of interest. Moreover, the hierarchical nature of energy flow implies that IVR out of a specific ZOBS will occur on several timescales. The shortest timescale coming from the first few tiers and the longest timescale representing the final, presumably irreversible, flow into a ``bath" of states in the final high density tiers. The dilution factor $\sigma_{\bf v}$ then yields the total number of zeroth-order states that participate in the IVR dynamics. 

Several experiments support the tier model and this perspective has recently been invoked to  understand vibrational energy transport in fairly large molecules\cite{Callegari1997,Kasyanenkoetal2011} as well as in molecule-surface reactions\cite{KilleleaUtz2013,bouakline19}. However, there are certain issues that need to be clarified. Firstly, how is the QNS explored? The value of the dilution factor is a long time  measure and it masks the potential non-exponential IVR dynamics. Secondly, the inherent assumption in the tier model for no ``backflow" from states in tier ${\cal T}_{k+1}$ to the states in tier ${\cal T}_{k}$ means that subtle correlations are left unaccounted for in the energy flow dynamics. Thirdly, how much of the tier model is classical? In other words, to what extent are the quantum interferences and coherences crucial to the predictions of the tier model.

\subsubsection{State space viewpoint and a threshold for quantum ergodicity}

A more illuminating way to think about IVR is the quantum state space approach\cite{mgrub99,mgrub00,GruebeleWolynes2004,semparithi06,srihariIVR2013,Leitner2015}. Apart from allowing for a clear connection to the classical phase space dynamics, as discussed later, the state space picture generalizes the tier model. In addition, the state space viewpoint directly highlights any anisotropy in the IVR dynamics. Several excellent reviews\cite{mgrub00,srihariIVR2013,Leitner2015} already exist for the state space approach to IVR and here we will briefly highlight the key predictions of the model within the LRMT perspective. Consider the time evolution of a ZOBS
\begin{equation}
    |{\bf v}(t)\rangle = e^{-i H t/\hbar} |{\bf v}\rangle \equiv \sum_{\alpha} C_{\alpha {\bf v}} e^{-i E_{\alpha} t/\hbar} |\alpha \rangle
    \label{zobstevolve}
\end{equation}
with $C_{\alpha {\bf v}} \equiv \langle \alpha|{\bf v} \rangle$ and $E_{\alpha}$ being the eigenvalues associated with the eigenstates. The central point, given the locality of the mode-mode couplings in molecules, is that $C_{\alpha {\bf v}}$ and $E_{\alpha}$ are correlated, once the $N_{\rm loc}(|{\bf v}\rangle)$ states coupled to the ZOBS get populated for $t \geq \tau_{\rm loc}$. As argued by Gruebele\cite{GruebeleBigwood1998}, in the absence of correlation the survival probability exhibits exponential decay
\begin{equation}
    P_{\bf v}(t) \sim (1 - \sigma_{\bf v}) e^{-t/\tau_{\rm GR}} + \sigma_{\bf v}
    \label{survexplaw}
\end{equation}
with $\tau_{\rm GR}$ being the golden rule timescale. The form above correctly accounts for the infinite time limit of the survival probability. However, correlated dynamics on the intermediate timescales $\tau_{\rm loc} < t < \tau_{\sigma}$ are expected to give rise to power law decays\cite{GruebelePNAS1998}
\begin{equation}
    P_{\bf v}(t) \sim (1 - \sigma_{\bf v}) \left[1 + \frac{2 t}{\tau D_{\bf v}} \right]^{-D_{\bf v}/2} + \sigma_{\bf v}
    \label{survpowlaw}
\end{equation}
with $D_{\bf v}$ being now interpreted as the effective dimensionality of the QNS associated with the IVR dynamics of the ZOBS. In the limit that the ZOBS completely explores the QNS, assuming a normal diffusive motion, one has the intuitive result $D_{\bf v} \approx (f-1)$. However, as shown by several experimental and theoretical studies\cite{GruebeleBSTR96,GruebelePNAS1998}, the IVR dynamics is much more interesting - even for large molecules and at substantial excitation energies $D_{\bf v} \ll (f-1)$. In other words, on the average, very few vibrational modes at any given time are involved in the IVR process. This, given the locality of bonding in molecules, is not entirely surprising. What is, however, significant is that it is not the rates but the power law exponent that is crucial for the IVR dynamics. An additional wrinkle on this comes from the fact that the assumption of normal diffusion in the QNS and the time-independence of $D_{\bf v}$ needs to be substantiated from careful dynamical studies. For example, it has been shown\cite{manikandankesh14} that $D_{\bf v}$ can change over time in the SCCl$_{2}$  molecule. Further examples and connections to the classical phase space transport are provided in sec.~\ref{Bunkerbelowdiss}.

\begin{figure}
\centering
  \includegraphics[width=1\linewidth]{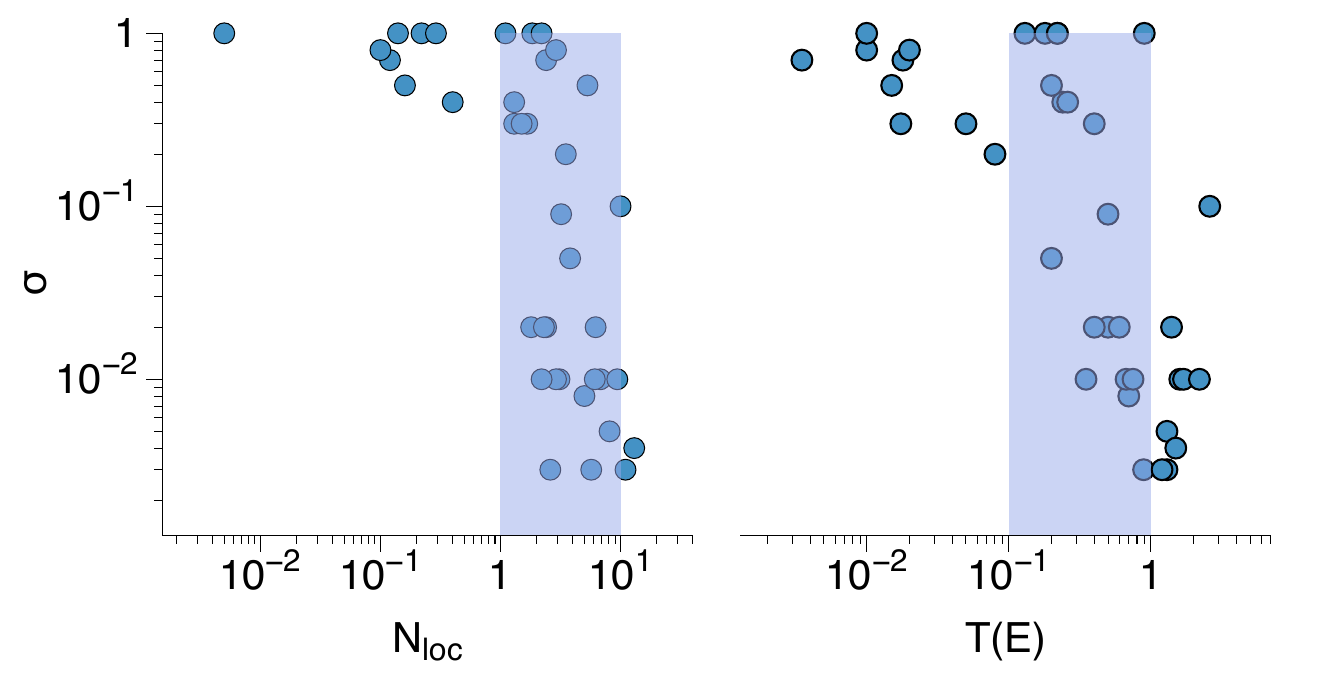}
  \caption{Dilution factor $\sigma$ versus number of local coupled states $N_{\rm loc}$ (left) and quantum ergodicity measure $T(E)$ (right) for the data\cite{macdonald83} of Stewart and McDonald corresponding to CH-stretch fundamentals of various molecules. The shaded portion in both plots indicate the sharp transition zones.}
  \label{fgr:sigmanlocT}
\end{figure}

The correlated intermediate time IVR dynamics and the associated power law decay can be elegantly rationalized based on the mapping of the IVR problem to the model of Anderson localization. Following the initial work by Logan and Wolynes\cite{loganwolynes1990}, several studies have established the utility of thinking of IVR as  transport in the QNS.  Indeed, the development of the LRMT model by Leitner and Wolynes\cite{leitnerwolynes1997} and the Bose-statistics triangle rule (BSTR) model\cite{GruebeleBSTR96,WongGruebele2001} by Gruebele and coworkers suggest that correlated IVR dynamics should be expected in general. In particular, a threshold for the onset of quantum ergodicity\cite{loganwolynes1990,LeitnerWolynesCPL1996,leitnerwolynes1997,Leitner2015} was put forward in terms of the measure
\begin{equation}
    T(E) = \frac{2 \pi}{3} \left[\sum_{\cal Q} \langle |V_{\cal Q}^{\rm eff}| \rangle K_{\cal Q} D_{\cal Q}(E) \right]^{2}
    \label{lrmt_qet}
\end{equation}
with $V_{\cal Q}^{\rm eff}$, $K_{\cal Q}$, and $D_{\cal Q}$ being the effective matrix element, number of connected states, and the local density of states respectively that are lying a distance ${\cal Q}$ away from a ZOBS of interest. The quantum ergodicity transition is then signalled by $T(E) \sim 1$. The corresponding limit is approached for $N_{\rm loc} \sim 10$ and  IVR is predicted to be very sensitive to specific anharmonic resonances for intermediate values of $N_{\rm loc}$. This is illustrated in Fig.~\ref{fgr:sigmanlocT} where the original experimental data\cite{macdonald83} of Stewart and McDonald on the dilution factors associated with the CH-stretch fundamentals of various molecules are plotted in terms of the LRMT parameters\cite{bigwgrubleitnwol98}. As noted by Bigwood et al\cite{bigwgrubleitnwol98}, the transition is not as sharp as seen in Fig.~\ref{fgr:sigmanlocT} when plotted against the total density of states.
Indeed, this fact can potentially be used to structurally modify molecules in order to decouple certain modes and modulate the IVR rates.
Further analysis by Leitner and Wolynes\cite{LeitnerWolynesCPL1996,Leitner2015} also led to an expression for the distribution of the dilution factors for $T(E) < 1$ as
\begin{equation}
    P(\sigma) = \frac{\gamma}{\sqrt{\sigma (1-\sigma)^{3}}} e^{-\pi \gamma^{2} \sigma/(1-\sigma)}
    \label{lrmt_dildist}
\end{equation}
where the quantity $\gamma \equiv [3 T(E)/(2 \pi (1-T(E))]^{1/2}$. Interestingly, eq.~\ref{lrmt_dildist} predicts a bimodal distribution for intermediate values of $T(E)$ and $N_{\rm loc} > 1$, indicating that even in the facile IVR regimes there are protected ZOBS. 
The predictions of LRMT have been tested for a number of molecules with considerable success\cite{bigwgrubleitnwol98}. Essentially, the threshold energy $E$ for which QET occurs relative to the conformational barrier heights or unimolecular dissociation energies determines the extent to which finite IVR timescales need to be accounted for with the possibility of the system exhibiting non-RRKM behaviour. For example, LRMT studies on the  isomerization dynamics of stilbene\cite{leitnerwolynesCPL97,Leitner2003}, cyclohexane ring inversion\cite{leitnerACP2005}, and the influence of torsional mode on the conformational dynamics of butanal\cite{leitgrubmolphys08} reveal deviations from RRKM rates ranging from one to several orders of magnitude. Several other examples for the application of LRMT can be found in the recent review by Leitner\cite{Leitner2015}.

\begin{figure}
\centering
  \includegraphics[width=0.75\linewidth]{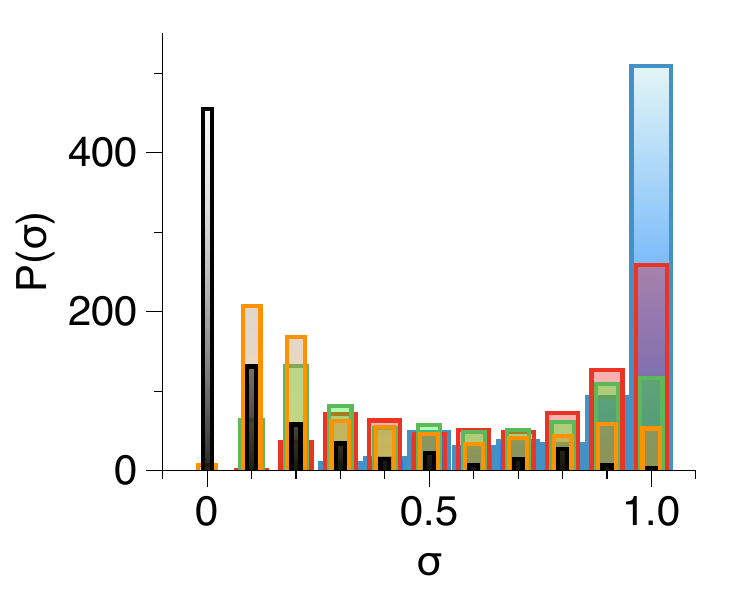}
  \caption{Dilution factor distribution for the model Hamiltonian eq.~\ref{ham_qm_ccm} with varying coupling strengths $(\beta_1,\beta_2,\beta_3) \times 10^{-5} \equiv [\beta_1,\beta_2,\beta_3] = [20,1,20] $ (blue), $[50,5,50] $ (red), $[100,10,100] $ (green), $[200,10,200]$ (orange), and $[500,50,500]$ (black). Note that the zeroth-order states  with energy $39.5 \leq E \leq 40.5$ are considered. The different cases are shown with differing histogram widths for clarity.}
  \label{fgr:df_hist}
\end{figure}

Note that the LRMT,  appropriate for large molecules with low average occupancy per oscillator, is statistical or probabilistic in  nature i.e., it assumes that the zeroth-order energies ${\cal E}_{j}$ and the couplings $\Phi_{\bf m}$ are random variables with appropriate, physically meaningful, average and variance. However, the local nature of the molecular couplings is built into the theory.    The higher order couplings (${\cal Q} > 3$)  are assumed, based on the elegant work\cite{BigwoodGruebeleCPL1995,MadsenGrub1997,Pearman1998} by Gruebele and coworkers, to scale as $V_{\cal Q} \sim \Phi_{3} \bar{a}^{3-{\cal Q}} M^{{\cal Q}/2}$ with $M$ being the average number of quanta per oscillator and $\bar{a}$ being a scale factor. This particular scaling implies that $V_{\cal Q}$ decrease exponentially  with increasing ${\cal Q}$ - in accordance with the 
intuitive expectation that low order resonances dominate over higher order resonances. Thus, combined with the scaling argument and explicit consideration of the dominant resonances, LRMT has proved to be a rather versatile theory. 

\begin{figure}
\centering
  \includegraphics[width=1\linewidth]{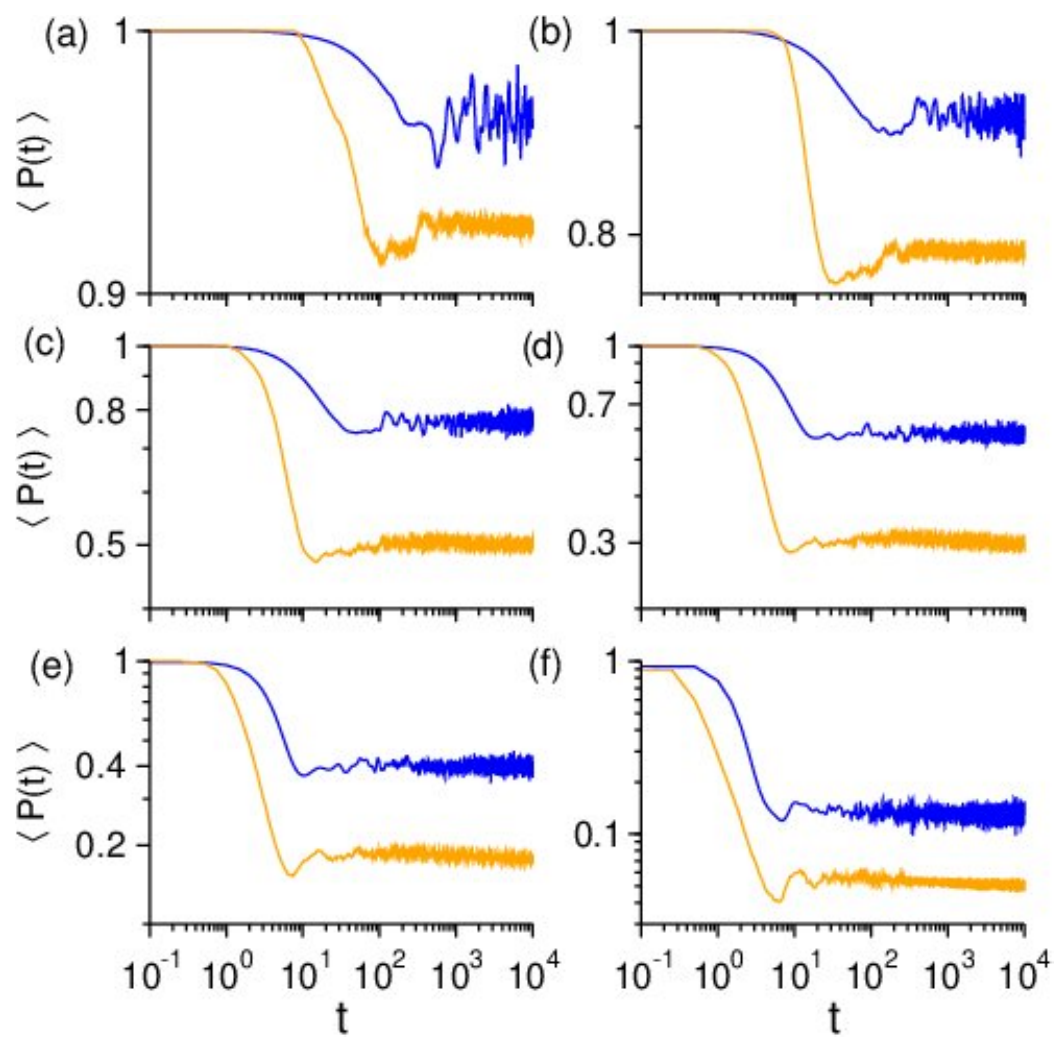}
  \caption{Microcanonically averaged quantum (blue) and classical (orange) survival probabilities for the model Hamiltonian eq.~\ref{ham_qm_ccm}  with coupling strengths (a) $[5,1,5]$, (b) $[20,1,20]$, (c) $[50,5,50]$, (d) $[100,10,100]$, (e) $[200,10,200]$, and (f) $[500,50,500]$. For classical probability, the phase space at $E=40$ is sampled uniformly. For quantum probability, the zeroth order states with energy $39.9 \le E \le 40.1$ are considered. Note that the notation for the coupling strengths is as in Fig.~\ref{fgr:df_hist} and will be followed for the rest of the figures showing results for the model Hamiltonian.}
  \label{fgr:micro_sp_cl_qm}
\end{figure}

\subsubsection{Need for a classical dynamical perspective on LRMT?}

As noted above, the LRMT and scaling theory\cite{schofieldwolynes1993,Leitner2015,srihariIVR2013} predictions are expected to hold for sufficiently large molecules since the probabilistic arguments\cite{loganwolynes1990} yielding  the localization-delocalization transition are sharp only in the thermodynamic limit. Nevertheless, an intriguing aspect that has emerged from the various investigations of  IVR dynamics in small molecules $(f \sim 3-5)$ is that the power law scalings, the bimodality of the dilution factor distributions, and other LRMT predictions seem to hold\cite{srihariCPL99,semparithi06,manikandankesh14}. For example, let us consider the model $f=3$ effective Hamiltonian introduced before (cf. eq.~\ref{ham_qm_ccm}). In Fig.~\ref{fgr:df_hist} we show the dilution factor distributions with increasing strengths of the anharmonic resonances. Interestingly, in accordance with the LRMT predictions of eq.~\ref{lrmt_dildist}, a clear bimodal behaviour is observed. Furthermore, in Fig.~\ref{fgr:micro_sp_cl_qm} the microcanonically averaged survival probability
\begin{equation}
    \langle P(t) \rangle = \frac{1}{N_{\Delta E}} \sum_{{\bf v} \in E+\Delta E} P_{\bf v}(t)
\end{equation}
at $E=40$ with $\Delta E = 0.1$, corresponding to $N_{\Delta E} = 151$ near-degenerate ZOBS, are shown for increasing resonant coupling strengths. The model $f=3$ system appears to show a behaviour which is very similar to the one observed by Schofield, Wyatt, and Wolynes in their quantum IVR studies on a highly degenerate $f=6$ coupled Morse oscillators system\cite{SchofieldWolynesWyatt1995}.  A clear intermediate time power law behaviour is observed. Although the observation of bimodality of $P(\sigma)$ and power law scaling of the survivals for the model is already interesting, the real ``puzzle" comes from the fact that the classical analog of $\langle P(t) \rangle$  also seem to exhibit power law scaling. In fact, beyond some values of the coupling strengths, the effective state space dimensionality seems to be nearly identical for the quantum and classical IVR dynamics. What is the reason for such an exquisite classical-quantum correspondence? Is it surprising or accidental that this model $f=3$ system seems to agree with the LRMT predictions?  

One reason may be that the central tenet of LRMT regarding the importance of local density of states implies that the smallness of the molecule is not a relevant issue - of course, the localization-delocalization transition will not be ``sharp". A second reason may have to do with the observation that the picture of IVR as diffusion in QNS has a natural classical counterpart, given the correspondence $J_{k} \leftrightarrow \hbar(n_{k} + 1/2)$, in terms of diffusion of zeroth-order actions $J_k$ in the classical phase space\cite{semparithi06,manikandankesh14,pittmanheller16}. Beyond this basic correspondence expectation,  several studies on relaxation in isolated quantum systems\cite{borgonovietal16,michailidisetal20}, in the context of ETH and MBL, have benefited from careful considerations of the underlying classical phase space dynamics. In the context of IVR, as one approaches the QET the dilution factors in eq.~\ref{lrmt_dildist} approach the Porter-Thomas distribution which, in terms of the Gaussian orthogonal ensemble in random matrix theory (RMT), signals a strongly chaotic classical phase space\cite{BohigasSchmit1984,WintgenFriedrich1987}. Several studies have also established the sensitivity of the averaged survival probability to the nature of the classical dynamics\cite{manikandankesh14,alusetal14,casatietal00}. For example, Wilkie and Brumer\cite{WilkieBrumerPRL1991} have shown that in the irregular case the average survival probability will fall below its long time average. More recently, for Hamiltonians whose underlying classical dynamics is chaotic but with local interactions, a thorough investigation of the various timescales associated with the average survival probability have been made\cite{SchiulazSantosPRB2019}. In particular, importance of the correlation hole and the so called Thouless time $\tau_{\rm Th}$, signalled by the minimum in $\langle P(t) \rangle$ has been emphasized\cite{SchiulazSantosPRB2019}. A third, and important, reason has to do with the fact that power law behaviour of the Poincar\'{e} recurrence statistics\cite{lange2016,dasbacker20} of an ensemble of orbits in the classical phase space has connections to the quantum survival probability. This raises the intriguing issue of whether one can interpret the effective IVR dimensionality $D_{\bf v}$ (cf. eq.~\ref{survpowlaw}) in the QNS with specific structures in the classical phase space. 

Thus, the observations above indicate that  a classical ``deconstruction" of the QET criterion in eq.~\ref{lrmt_qet} may yield deeper insights into the IVR dynamics.  However, making progress in this direction requires us to understand and characterize the classical phase space transport in systems with $f \geq 3$.  There are several reasons for this. Firstly, and as noted above, it is exceedingly difficult to drive realistic molecular systems into regimes of hard chaos wherein RMT arguments may be usefully invoked. Thus, typical molecular phase spaces at physically relevant levels of excitations are expected to display a mixture of regular and chaotic dynamics.  In this context, as explained below, systems with $f \geq 3$ are fundamentally diferent from $f=2$. Secondly, in the scaling theory approach of Schofield, Logan, and Wolynes\cite{schofieldwolynes1993}, the distinction between critical $(\sim t^{-1}$) and diffusive $(\sim t^{-(f-1)/2}$) regimes of the average survival probability cannot be made even for $f=3$. 
Thirdly, although LRMT and scaling approaches identify the regimes of facile IVR and allow for corrections to the RRKM rates, one cannot directly infer the crucial degrees of freedom that are involved in the IVR mechanism. Clearly, this information is paramount for identifying and modifying the IVR pathways i.e., control; the extent to which there is a correspondence between the classical and quantum IVR pathways should prove to be a valuable guide to designing rational control fields.  

\section{Classical dynamical viewpoint on IVR}
\label{CMIVR}

We start with a dimensionality count of the relevant objects in phase space. The discussion below is limited to conservative Hamiltonian systems. For a $f$ degrees of freedom system described by a Hamiltonian $H({\bf P},{\bf Q})$ the phase space $({\bf P},{\bf Q})$ is $2f$-dimensional and the constant energy ``surface" $H({\bf P},{\bf Q})=E$ is $(2f-1)$-dimensional. The Hamilton's equations of motion
\begin{eqnarray}
\dot{Q}_{k} &=& \frac{\partial H({\bf P},{\bf Q})}{\partial P_{k}} \nonumber \\
\dot{P}_{k} &=& - \frac{\partial H({\bf P},{\bf Q})}{\partial Q_{k}}
\label{hameq}
\end{eqnarray}
for $k=1,\ldots,f$ yield the time evolution (flow) of a specific initial point in the phase space. In general, the resulting trajectory can be regular or irregular. In the former case the trajectory is confined to a $f$-dimensional torus and in the latter case the trajectory, if chaotic, explores the entire constant energy shell. For $f=2$ the  three dimensional energy shell is divided by the regular two-dimensional tori. However, for $f \geq 3$ the regular tori no longer divide the constant energy shell and this geometrical fact is responsible for the significant difference between the transport in phase spaces with $f=2$ and $f \geq 3$.
Before bringing out this fundamental difference and its relevance to IVR in more detail, note that  the dimensional counting is far from a trivial exercise - such dimensionality arguments play a crucial role in identifying the appropriate invariant phase space structures that are relevant to chemical reaction dynamics. For example, in TST the so called normally hyperbolic invariant manifold (NHIM)\cite{wiggetalprl01,WaalkensTST2007,Uzeretal2002} has a dimensionality of $(2f-3)$ and forms the equator of the $(2f-2)$-dimensional dividing surface that divides the phase space into the reactant and product regions. The NHIM is an invariant manifold and hence cannot be crossed by trajectories, while the dividing surface can be. In fact, away from the TS, the stable and unstable manifolds attached to the NHIM, which itself is a dynamical object, are expected to be relevant for the IVR in the reactant and product regions.   This aspect, however, is yet to be investigated and will not be addressed here.

The central objects of interest are the various critical points of the flow in eq.~\ref{hameq}, as these points  are the pivots about which much of the insights can be obtained into the transport in phase space. For instance, in the context of TST, the relevant fixed points are those that are associated with rank-$k$ saddles i.e., regions in phase space associated with $k$ unstable (hyperbolic) directions and $(f-k)$ stable (elliptic) directions. Thus, $k=1$ is the most common case pertaining to the transition states. Expansion of the Hamiltonian about such points is properly done using the classical theory of normal forms\cite{WaalkensTST2007}. In the case of IVR studies, such expansions are usually done about the equilibrium points. The resulting normal form Hamiltonian is the classical analog of eq.~\ref{dunham}.

At this stage one can consider an initial phase space density $\rho({\bf P},{\bf Q},t=0)$ and study its time evolution under the classical flow generated by the Hamiltonian of interest. The rate at which the ensemble of initial conditions explores the constant energy shell is then intimately connected to the issue of ergodicity. Clearly, IVR is implicitly present in the time evolution of the phase space density, particularly if this is chosen appropriately. For instance, one can choose to work with the Wigner phase space density $W_{\bf v}({\bf P},{\bf Q}, t=0)$ associated with the ZOBS $|{\bf v}\rangle$. Nevertheless, despite the usual semiclassical advantages, ``dissecting" the Wigner density in terms of the structures in multidimensional phase space and identifying the dominant IVR pathways has hardly been attempted till date.  Instead, we focus here on analyzing the Hamiltonian in eq.~\ref{gen_ham} in terms of the general phase space features and identifying the role of the various nonlinear resonances - classical analogs of quantum anharmonic resonances. Such an analysis is useful in order to better understand the viewpoint of IVR as diffusion in the QNS, and essential for any attempt which seeks to establish the classical analog, if it exists, of QET.

\begin{figure*}
\begin{center}
\includegraphics[width=0.75\linewidth]{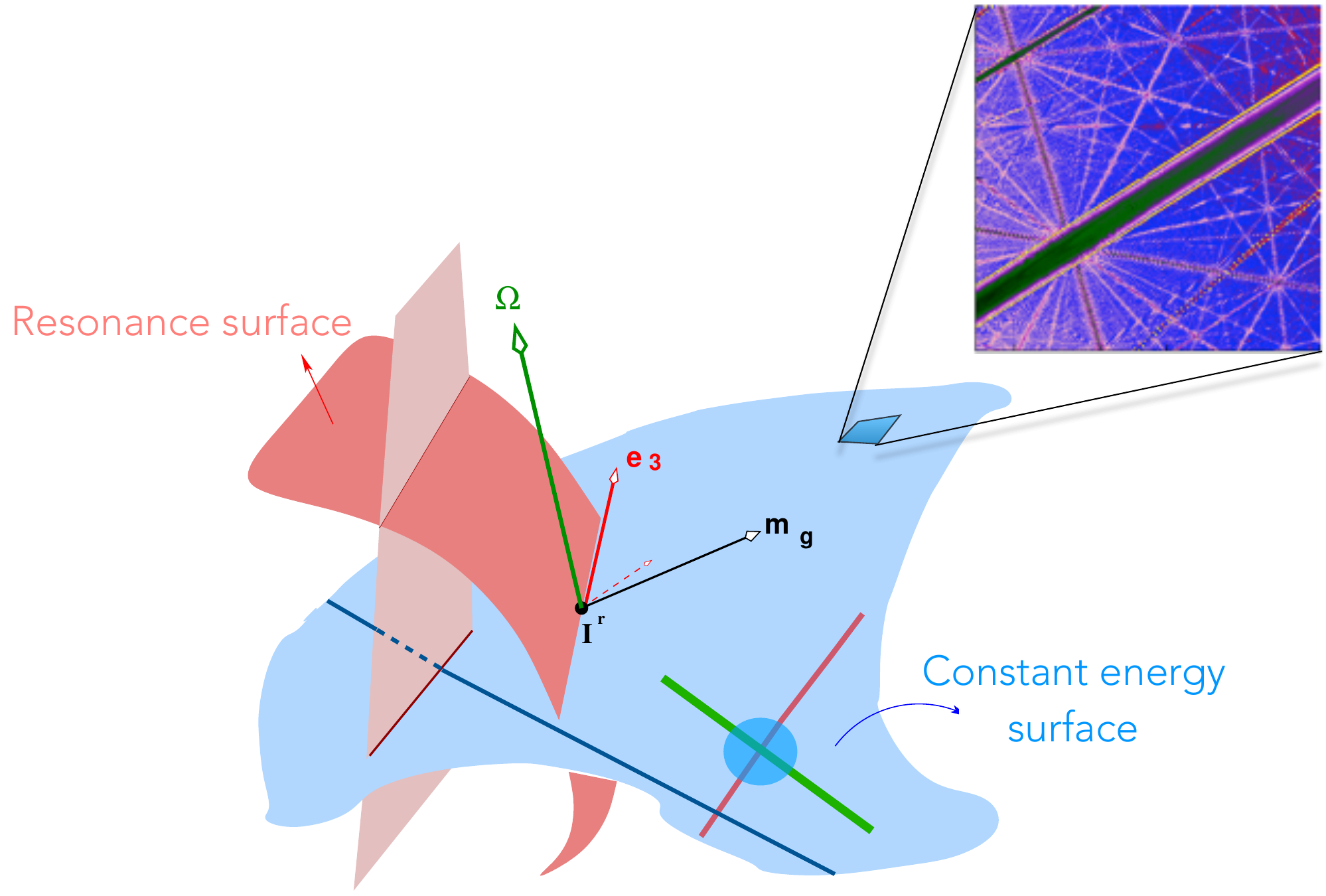}
\end{center}
\caption{A sketch illustrating the geometry of Arnold web for $f=3$. Two example resonance planes intersecting the constant energy surface (CES) are shown. Intersection of other resonance planes are shown as lines of different thickness on the CES. A resonance junction is highlighted with a blue circle. Note that, typically, the resonances are dense on the CES as shown in the example zoom plot. The zoomed figure corresponds to an actual computation for one of the models of Bunker\cite{Bunker1,SKPKYKS2020}.}
\label{fgr:webgeom}
\end{figure*}

Consider the Hamiltonian eq.~\ref{gen_ham} with, as stated before, the zeroth-order part $H_{0}({\bf P},{\bf Q})$ representing a set of uncoupled oscillators. Note that $H_{0}({\bf P},{\bf Q})$ can be approximated by harmonic oscillators with frequencies $\omega_{k}$ ($k=1,\ldots,f$) for low energies. However, depending on the molecular size and extent of anharmonicities, at higher energies it is more appropriate to think of the zeroth-order part as a set of anharmonic oscillators. In classical dynamics the oscillators are nonlinear since their frequencies $\Omega_{k}(E_{k})$ depend on the energy $E_{k}$. Analogously, in quantum mechanics one has an equivalent distinction. For harmonic oscillators the energy level spacings are constant whereas for anharmonic oscillators, like the Morse oscillator for example, the energy level spacings depend on the vibrational level. Classically, the dynamics of $H_{0}({\bf P},{\bf Q})$ is integrable and motion lies on invariant $f$-dimensional tori, parametrized by angles ${\bm \theta}$ with $\dot{\theta}_{k} = \Omega_{k}$. The addition of the coupling terms, in general, destroys the integrability with the possibility of irregular motion. Nevertheless, the celebrated Kolmogorov-Arnold-Moser (KAM) theorem assures us that for sufficiently small coupling strengths most of the tori are only distorted and do not break up. According to KAM there are, however, tori that will break up even for the slightest perturbation. These tori are the ones which fail to satisfy the so called Diophantine condition i.e.,
\begin{equation}
    |{\bf r} \cdot {\bm \Omega}| = |\sum_{j=1}^{f} r_{j} \Omega_{j}| \geq \frac{\gamma}{||{\bf r}||^{\tau}}
    \label{diophant}
\end{equation}
with $\gamma, \tau$ being some positive constants and ${\bf r} = (r_{1},r_{2},\ldots,r_{f}) \in {\mathbb Z}^{f}\setminus \{0\}$ being an integer vector, said to be of order $||{\bf r}|| \equiv |r_{1}| + \ldots + |r_{f}|$.  Note that the failure of the condition eq.~\ref{diophant} is essentially indicating that any attempt to perturbatively transform away the mode-mode couplings will not succeed. In particular, for ${\bf r} \cdot {\bm \Omega} \approx 0$ one has a nonlinear resonance - precisely the condition that signals active IVR involving the various vibrational modes! This so called ``small-denominator" problem, just as in quantum CVPT, implies that an accurate description of the dynamics cannot be obtained by an integrable normal form and one has to work with resonant normal forms. Quantum mechanically, this is equivalent to the fact that the Dunham expansion eq.~\ref{dunham} is insufficient and one needs to include the various anharmonic resonances. The type (order) and number of resonances depends on the total energy of interest. The various effective or spectroscopic Hamiltonians, including the model in eq.~\ref{ham_qm_ccm}, that have been proposed for a number of molecules are nothing but resonant normal forms. 

\subsection{Classical limit Hamiltonian}

The classical limit Hamiltonian corresponding to the resonant normal form quantum Hamiltonian can be obtained via the Heisenberg correspondence
\begin{equation}
    {\bf a} \Longleftrightarrow \sqrt{{\bf J}} e^{-i {\bm \theta}} \,\,\,\,\, ; \,\,\,\,\, {\bf a}^{\dagger} \Longleftrightarrow \sqrt{{\bf J}} e^{i {\bm \theta}} 
    \label{heisencorr}
\end{equation}
where $({\bf J},{\bm \theta})$ are the action-angle variables associated with the zeroth-order Hamiltonian in eq.~\ref{gen_ham}, assumed to be an integrable Hamiltonian. Note that such an assumption is valid since we consider the zeroth-order Hamiltonian to be a sum of $f$ independent anharmonic oscillators. It is also worth emphasizing that an explicit canonical transformation $({\bf P},{\bf Q}) \rightarrow ({\bf J},{\bm \theta})$ may not be readily available in general, except when dealing with normal modes (harmonic oscillators) or local modes (Morse oscillators, for example). Nevertheless, the integrability of $H_{0}({\bf P},{\bf Q})$ ensures the existence\footnote[3]{We ignore here the possibility of there being obstructions to defining global action-angle variables i.e., the phenomenon of monodromy\cite{duistermaat}. Although monodromy, and its spectral manifestations\cite{childetal99,cushmanetal04,zobovetal05,sadovskiietal06,dullinetal18}, are strictly relevant only in completely integrable systems, the effect does survive weak integrability-breaking perturbations. However, monodromy effects may not be very relevant for the regimes of IVR we are interested in.} of $f$ action variables which mutually commute in the Poisson bracket sense i.e., $\{J_{i},J_{j}\}=0$ for $i,j = 1,2,\ldots,f$. Moreover,  the action-angle representation, apart from providing a natural representation to connect to the QNS, are ideal for understanding the stability properties of phase space initial conditions and describing the geometry of the connected network of the anharmonic resonances which mediate the IVR process. 

The transformation eq.~\ref{heisencorr} yields the classical analog of the Hamiltonian in eq.~\ref{hamcompact} as
\begin{eqnarray}
    H({\bf J},{\bm \theta}) &=& H_{0}({\bf J}) + \sum_{\bf m} \Phi_{\bf m} \prod_{j=1}^{f} \left(2\sqrt{ J_{j}} \cos \theta_{j} \right)^{m_{j}} \nonumber \\
    &\equiv& H_{0}({\bf J}) + \epsilon \sum_{\bf r} f_{\bf r}({\bf J}) \cos({\bf r} \cdot {\bm \theta}) \label{ham_actangrep}
\end{eqnarray}
with ${\bf r} = (r_{1},r_{2},\ldots,r_{f}) \in {\mathbb Z}^{f}\setminus \{0\}$ i.e. an integer vector excluding the null vector. The factor $\epsilon$ has been introduced as a bookkeeping term and $f_{\bf r}({\bf J})$, related to the coupling constants $\Phi_{\bf m}$, are functions of the zeroth-order actions. For $\epsilon=0$ we have an integrable system and the actions are conserved quantities. The system is called as near-integrable for sufficiently small $\epsilon \ll 1$, whereas for larger values of $\epsilon$ one has a nonintegrable system with a phase space composed of both regular and chaotic dynamics. The key terms for IVR in eq.~\ref{ham_actangrep} come from  the slow variation of certain angle variables i.e., $\dot{\psi}_{\bf r} \equiv d({\bf r} \cdot {\bm \theta})/dt \approx 0$. Such a condition implies ${\bf r} \cdot {\bm \Omega} \approx 0$, leading to a nonlinear resonance due to the commensurability between certain mode frequencies. 

In order to understand the ``geography" of the resonances consider the zeroth-order frequencies ${\bm \Omega}_{0}({\bf J}) \equiv \nabla_{{\bf J}} H_{0}$.  A resonance condition ${\bf r} \cdot {\bm \Omega}_{0}({\bf J}) \approx 0$ among the different frequencies, satisfied for resonant actions ${\bf J}^{r}$, implies the condition ${\cal R}_{\bf r}({\bf J}^{r}) = 0$
representing a $(f-1)$ dimensional surface in the zeroth-order action space.  A sketch of the geometry is shown in Fig.~\ref{fgr:webgeom} for the $f=3$ case. Each independent resonance is thus represented by an appropriate surface. A resonant surface intersects the  $(f-1)$ dimensional constant zeroth-order energy surface $H_{0}({\bf J}) = E_{0}$ forming a $(f-2)$-dimensional surface. The collection of all such $(f-2)$-dimensional surfaces forms an intricate connected network called as the Arnold web. Note that the web is connected only for $f \geq 3$. For finite coupling $\epsilon \neq 0$, the resonance surfaces gain widths proportional to $\sqrt{\epsilon}$ and their order.  

As indicated in the sketch in Fig.~\ref{fgr:webgeom}, a specific resonance may or may not intersect the constant energy surface. Moreover, intersection of independent resonances to form junctions is also dependent on the energy of interest. Nevertheless, on a given energy surface the resonances form a dense, interconnected network. One such example is shown in Fig.~\ref{fgr:webgeom}, representing a enlargement of a portion of a computed Arnold web. This is due to the fact that in $f \geq 3$ existence of even one junction implies the existence of an infinity of resonances! Thus, if two independent resonances ${\bf r} \cdot {\bm \Omega}_{0}({\bf J}) \approx 0$ and ${\bf r}' \cdot {\bm \Omega}_{0}({\bf J}) \approx 0$ then 
\begin{equation}
    l ({\bf r} \cdot {\bm \Omega}_{0}({\bf J})) + m ({\bf r}' \cdot {\bm \Omega}_{0}({\bf J})) \approx 0
\end{equation}
for all integers $(l,m) \neq (0,0)$. How does one even attempt to describe the IVR in the presence of such complexity? The answer lies within a powerful approach introduced by Nekhoroshev\cite{Nekhoroshev1977} nearly half a century ago. Accordingly, one considers all resonances up to a finite order $K$ and neglects all other higher order resonances. Following Nekhoroshev one can now define three domains on the Arnold web
\begin{enumerate}
    \item No-resonance domain: These correspond to points on the web that are ``far enough" from all the resonances. The dynamics is nearly integrable in such domains with the frequencies ${\bm \Omega}_{0}$ remaining unchanged.
    \item Single resonance domain: These are regions which are influenced by only one resonance of order less than $K$. The dynamics is again nearly integrable, but the frequencies are no longer constant and can change in a direction (usually called as the fast-drift direction) that is transverse to the resonance zone. There is also the possibility of diffusion along a resonance (as indicated by the direction labeled ${\bf e}_{3}$ in Fig.~\ref{fgr:webgeom}), known as Arnold diffusion. However, the Arnold diffusion timescale is expected to be way too long for any relevant molecular consequences and will not be discussed further here.
    \item Double resonance domain: Here the point on the web is in the vicinity of a junction and, hence, under the influence of at least two independent resonances of order less than $K$. The dynamics is now non-integrable, leading to chaotic motion which, however, is bounded.
\end{enumerate}
From a molecular perspective the different domains correspond to the different types of IVR dynamics that one can expect from a ZOBS. The first case corresponds to no IVR, the second to a regular IVR with a polyad as in a single stretch-bend resonance, and the third case to irregular IVR with, as we shall see below, approximate polyads. Note that in the last case the IVR can exhibit a range of behaviour depending on the order and the number of independent resonances that form the junction. 

Given the set up as above, the celebrated Nekhoroshev's theorem provides results regarding the  stability of various types of initial conditions on the Arnold web. Roughly speaking\footnote[4]{Note that the precise statement of the theorem requires several technical definitions. We do not go into them here for two main reasons. Firstly, the technicalities, while very relevant, are beyond the scope of this Perspective, and our intention here is to convey the ``spirit" of the theorem. Secondly, several excellent accounts\cite{morbguzzo1996,Efthymiopoulos2008,Cincottaetal2014} of the Nekhoroshev theorem can be found in the literature that are amenable to non-mathematicians.}, for $\epsilon < 1$, an initial condition $({\bf J}(0),{\bm \theta}(0))$ satisfies
\begin{equation}
    ||{\bf J}(t)-{\bf J}(0)|| \leq \epsilon^{1/2(f-m)}
\end{equation}
for times
\begin{equation}
    |t| \leq \exp \left(c \epsilon^{-1/2(f-m)} \right)
\end{equation}
with $c$ being a positive constant and $m$ being the multiplicity of the domain. Thus, for single resonance domains $m=1$, whereas for double resonance domains $m=2$. In other words, initial conditions near junctions are more stable as compared to being near single resonances. 

A key point that is worth emphasizing is that the division into the various domains above is a function of the coupling $\epsilon$, and there is a delicate interplay between the timescales of interest and the maximal order $K$ that one should consider. Interestingly, similar considerations are typically invoked during a CVPT analysis of molecular Hamiltonians. Additionally, a caveat needs to be stressed at this stage - the formal proof of Nekhoroshev's theorem involves bounds that would put several cases in the literature strictly out of the Nekhoroshev regime. Nevertheless, as pointed out by Morbidelli and Froeschl\'{e}\cite{morbfroes96}, there may be some ``leniency" when it comes to the numerical investigations on physical systems. In any case, a relevant limit corresponds to large coupling strengths for which the above domains cannot be strictly defined. One then is in the so called Chirikov regime\cite{Chirikov79} wherein, from the molecular perspective, the IVR dynamics is expected to be highly irregular and faster than typical vibrational timescales. The transition from KAM, to Nekhoroshev, and finally to the Chirikov regime can be seen, for example, in the original work\cite{FroeschleScience2000} of Guzzo et al. The transition, and the connection to LRMT discussed in detail below, can be observed in Fig.~\ref{fgr:web_df_nloc}A as well. 

Although we do not discuss the various techniques for constructing the Arnold web, all the webs shown here utilize the fast Lyapunov indicator (FLI) method. A basic introduction is given in the Appendix and for details we refer the reader to the original literature\cite{Skokos2016,Froeschle2002}. Moreover, all the webs shown in this work are for a specific fixed choice of the angle values and are indicated in the figures. For near-integrable regimes different angle slices will more or less reveal the same features. However, for strong coupling strengths the web features do depend on the choice of the angle slice\cite{SKPKYKS2020}. Although we do not fully explore the different Arnold web slices in this work, suffice it to say that the qualitative arguments presented here are not affected. 

\section{Classical-quantum correspondence}
\label{three_resmodel}

We now illustrate the exquisite classical-correspondence between the QNS and the classical phase space for the model system in eq.~\ref{ham_qm_ccm}. We refer the reader to the recent work\cite{KarmakarKesh2018} on the same model system for the computational details and further examples and discussions on the competition between classical and quantum IVR pathways.  Similar analysis can be performed on any effective Hamiltonian that arises either from a CVPT calculation or by fitting the experimental spectra. In particular, the phase space structures highlighted in this perspective are expected to be generic and relevant for any effective Hamiltonian. Notably, in early studies\cite{atkinsloganpla92,atkinsloganjcp92} on a model $f=3$ system, Atkins and Logan have already pointed out the classical-quantum correspondence and made many, remarkably prescient, observations related to the resonance junctions, Lyapunov exponents, possibility of phase space sliced surface of sections, and the lack of Arnold diffusion!

\begin{table}
\centering
  \caption{ Parameters for the model three-resonance Hamiltonian}
  \label{tbl:example}
    \begin{threeparttable}

  \begin{tabular}{cc}
    \hline 
    $\omega_1$ & $1.1$ \\
    $\omega_2$ & $1.7$ \\ \vspace{2mm}
    $\omega_3$ & $0.9$ \\ 
    $\alpha_1$ & $-0.0125$ \\
    $\alpha_2$ & $-0.02$ \\ \vspace{2mm}
    $\alpha_3$ & $-0.0085$ \\ 
    $D_1$      & $48.4$ \\ 
    $D_2$      & $78.2$ \\
    $D_3$      & $47.6$ \\
    \hline
  \end{tabular}
  
\end{threeparttable}

\end{table}

\begin{figure*}
\begin{center}
\includegraphics[width=1\textwidth]{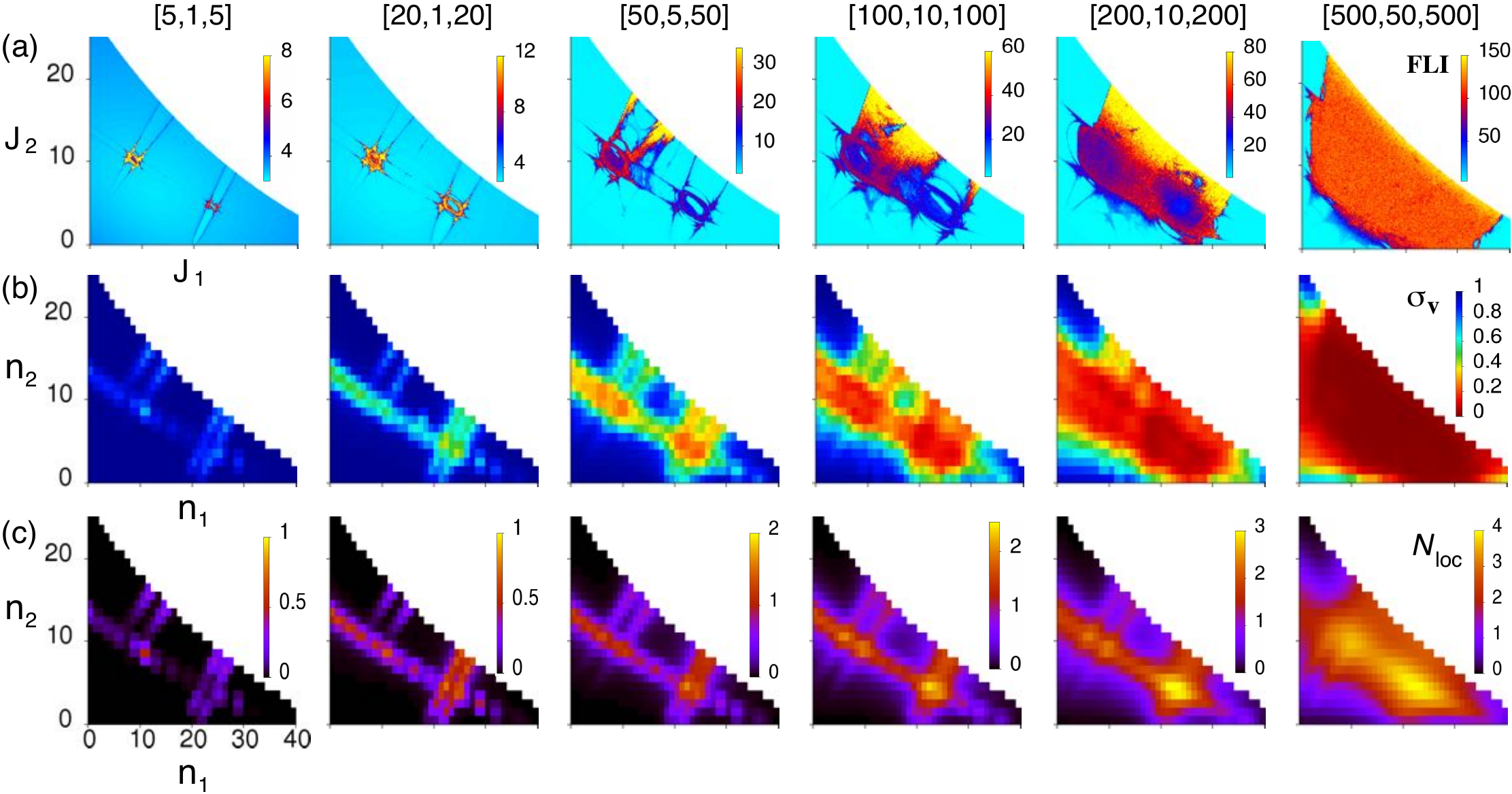}
\end{center}
\caption{The Arnold web (Panel (a)), corresponding dilution factor plots (Panel (b)) and $N_{\rm loc}$ plots (Panel (c)) for different couplings. Note that for dilution factor plots, the color bar is same for all the couplings. For the Arnold web computation, an uniform $500 \times 500$ grid is constructed in the $J_1-J_2$ plane with a final integration time  $t_f = 40$ and the total energy is $E=40$. The webs are shown for the angle slice $(\theta_1,\theta_2,\theta_3)=(\pi/2,\pi/2,\pi/2)$.For the quantum measures plots, the average $\sigma$ or $N_{\rm loc}$ is projected on $n_1-n_2$. The zeroth order states with energy $39.5 \le E \le 40.5$ are considered. See the text for details.}
\label{fgr:web_df_nloc}
\end{figure*}

Using eq.~\ref{heisencorr} the model $f=3$ Hamiltonian in eq.~\ref{ham_qm_ccm} has the classical analog\cite{ccmjstat1992,KarmakarKesh2018} ${\cal H}_{\rm model}({\bf J},{\bm \theta}) = {\cal H}_{0}({\bf J}) + V({\bf J},{\bm \theta})$ with
\begin{eqnarray}
 {\cal H}_{0}(\bf{J}) &=& \sum_{i=1}^{3} \left[\omega_i J_i + \frac{1}{2} \alpha_i J_i^2 \right] \\
  V({\bf J},{\bm \theta}) &=& 2 \sum_{k=1}^{3} \beta_k f_{{\bf r}_{k}}({\bf J}) \cos({\bf r}_{k} \cdot {\bm \theta}) \label{3-resham}
\end{eqnarray}
The three  nonlinear resonances ${\bf r}_{1}=(2,-1,0)$, ${\bf r}_{2} = (3,-2,0)$, and ${\bf r}_{3}=(0,1,-2)$ with coefficients $f_{{\bf r}_{1}}({\bf J}) = J_{1} J_{2}^{1/2}$, $f_{{\bf r}_{2}}({\bf J})=J_{2} J_{1}^{3/2}$, and $f_{{\bf r}_{3}}({\bf J})=J_{3} J_{2}^{1/2}$ are independent. For this model the resonances are $2$-dimensional planes, given that the nonlinear frequencies are linear in the actions, that may intersect the $2$-dimensional constant energy surface ${\cal H}_{0}({\bf J}) = E$ at lines. For instance, the plane representing the ${\bf r}_{1}$ resonance is given by
\begin{equation}
    2 \alpha_{1} J_{1} - \alpha_{2} J_{2} + (2 \omega_{1}-\omega_{2}) = 0
\end{equation}
and one can check that the above condition is essentially the same as the condition for the quantum degeneracy of the ZOBS $|v_{1},v_{2},v_{3} \rangle$ and $|v_{1} \pm 2,v_{2} \mp 1,v_{3} \rangle$. Thus, for $\beta_{1} \neq 0$ the near-degenerate ZOBS on the resonance plane are expected to couple strongly leading to IVR. With similar observations for the other two resonances in eq.~\ref{3-resham}, several questions arise. For instance 
\begin{itemize}
    \item What is the structure of the Arnold web as a function of the coupling constants? Are there special features on the web?
    \item Given that a point in the zeroth-order action space represents a potential ZOBS that can be accessed experimentally, does the quantum IVR dynamics respect the classical Arnold web structure? In other words, are the quantum measures of IVR such as the dilution factors $\sigma_{\bf v}$  and the local density of coupled states $N_{\rm loc}$ sensitive to the Arnold web?
    \item Is there any correlation between the nature of the web and the predictions of the LRMT?
\end{itemize}
In order to address these questions, in Fig.~\ref{fgr:web_df_nloc} we present the results of our computations on the model system at the total energy $E=40$ as a representative case. Similar observations hold for other energies below the dissociation threshold and for several other choices of the Hamiltonian parameters. Figure~\ref{fgr:web_df_nloc}(a) shows the Arnold web structure as a function of increasing resonance coupling strengths. Note that the strength of the fifth order ${\bf r}_{2}$ resonance is always kept, as expected in realistic systems, about an order of magnitude smaller than the remaining two resonances of third order. For low couplings the system is approximately in the Nekhoroshev regime and one can clearly see the three resonances with their appropriate widths. More importantly, two resonance junctions formed by the intersection of independent resonances, along with their local chaotic behaviour, can be seen. With increasing coupling strengths, the resonance widths increase and the system transitions to the Chirikov regime with considerable overlap of the resonances leading to large scale chaotic dynamics. In Fig.~\ref{fgr:web_df_nloc}(b) and (c) the QNS perspective is shown in terms of two different quantum IVR measures. The correspondence between the classical Arnold web structure in Fig.~\ref{fgr:web_df_nloc}(a)  and the  dilution factor measure (long time) in Fig.~\ref{fgr:web_df_nloc}(b) and number of locally coupled states (relatively shorter time) in Fig.~\ref{fgr:web_df_nloc}(c) is indeed striking. Specifically, the bimodal distribution of the dilution factors shown in Fig.~\ref{fgr:df_hist} can now be correlated with the Arnold web. The more local measure $N_{\rm loc}$, shown in Fig.~\ref{fgr:web_df_nloc}(c), is clearly reflecting the Arnold web structure. However, even for the largest coupling considered here it is apparent that there is heterogeneity in the QNS as compared to the web. More interestingly, $N_{\rm loc}$ much larger than unity are correlated with the two junctions on the Arnold web. According to LRMT the QET transition is expected to typically occur for $N_{\rm loc} \sim 5-10$. Combined with the fact that LRMT is applicable for molecules with several degrees of freedom wherein resonance junctions of various multiplicities abound, there are reasons to expect that the dynamics near the junctions will play a crucial role in approaching the QET. 

\subsection{Classical versus quantum IVR pathways}

Although Fig.~\ref{fgr:web_df_nloc} shows the close correspondence between QNS and the Arnold web, whether the dominant IVR pathways from a ZOBS of interest are similar in the quantum and classical cases is far from obvious. Similar effective state space dimensionalities do not necessarily imply similar IVR pathways! Clearly, this is an important issue for controlling the IVR dynamics of an initial state of interest with external fields. If indeed the dominant pathways are similar then control fields can be designed based on the classical phase space. On the other hand, if the IVR pathways have crucial differences then attempts to control the quantum IVR dynamics using classically inspired control fields is bound to fail. From Fig.~\ref{fgr:web_df_nloc} it is clear that the nature of the IVR dynamics is largely determined by whether the ZOBS of interest is located near a junction, near a single resonance, or away from resonances. We now illustrate that the location of a ZOBS in the QNS (or Arnold web) gives rise to IVR regimes with differing extent of   competition between purely quantum and classical IVR pathways. 

\begin{figure}
\centering
  \includegraphics[width=0.75\linewidth]{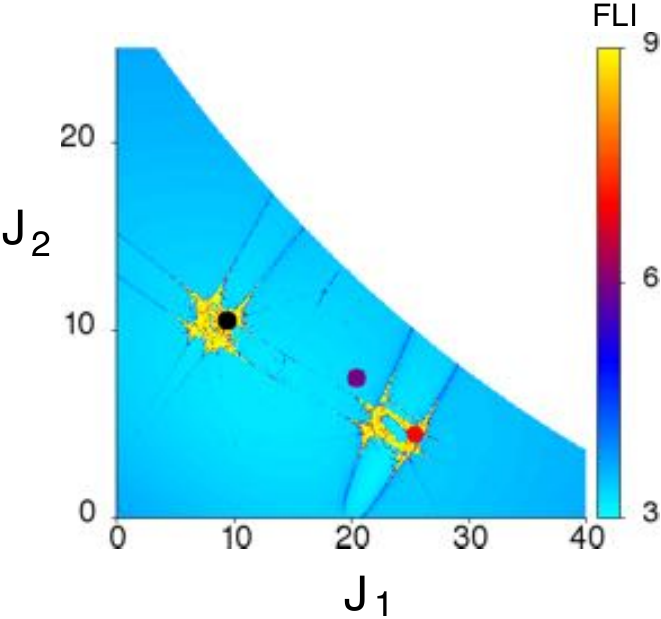}
  \caption{Locations of the three ZOBS on the web (for $[\beta_1, \beta_2, \beta_3] = [20,1,20]$): $|9,10,16\rangle$ (black dot), $|20,7,9\rangle$ (purple dot) and $|25,4,9\rangle$ (red dot). The zeroth order energy, $E^{0}_{\mathbf v}$, of these states are $40.33$, $40.28$ and $39.6$ respectively. The webs are shown for the angle slice $(\theta_1,\theta_2,\theta_3)=(\pi/2,\pi/2,\pi/2)$. See the text for details.}
  \label{fgr:states_location}
\end{figure}

\begin{figure}[htbp]
\begin{center}
\includegraphics[width=1.0\linewidth]{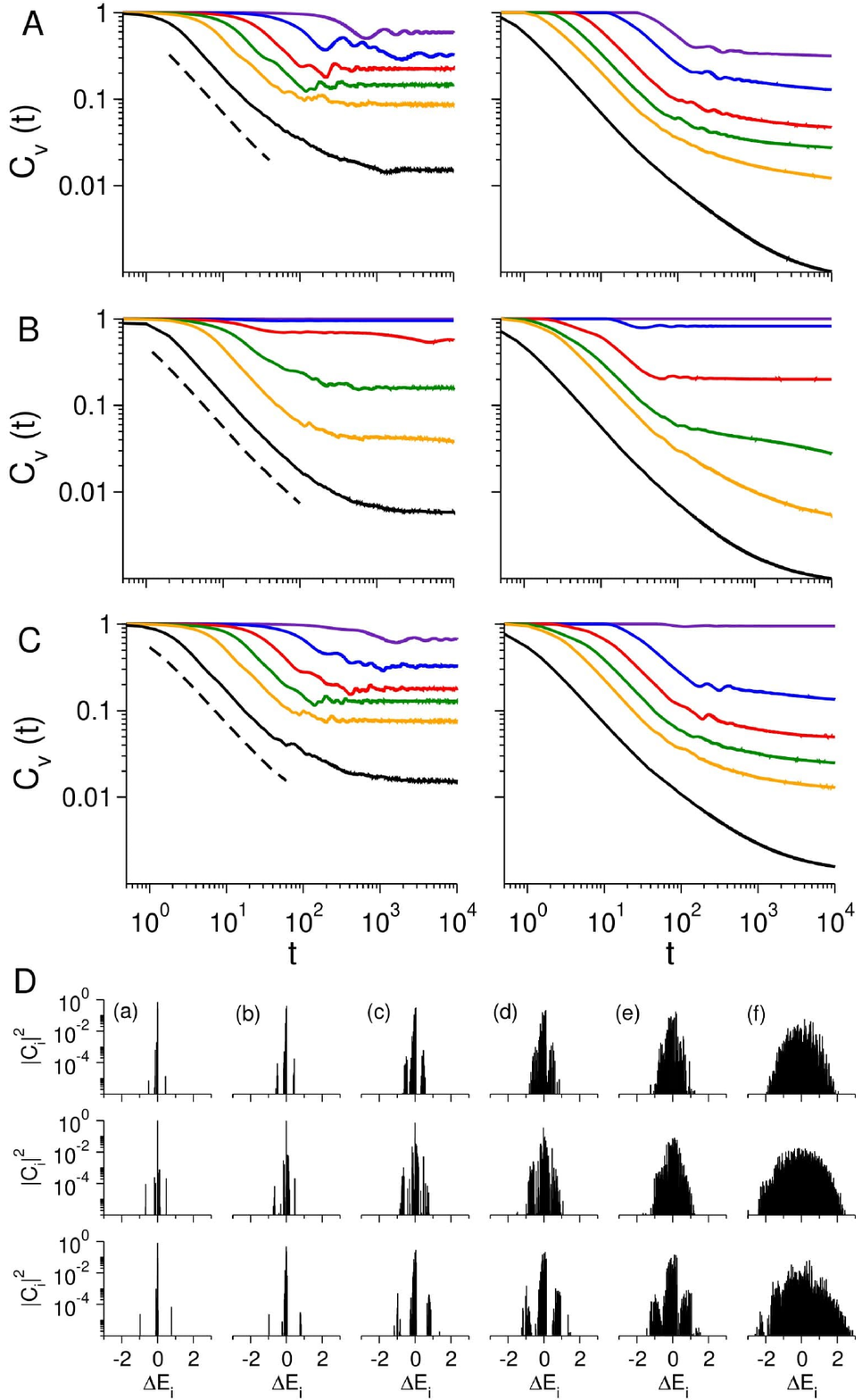}
\end{center}
\caption{Quantum (left) and classical (right) time-smoothed survival probabilities of the three ZOBS (A) $|9,10,16\rangle$, (B) $|20,7,9\rangle$, and (C) $|25,4,9\rangle$ at different couplings, $[5,1,5]$ (purple), $[20,1,20]$ (blue), $[50,5,50]$ (red), $[100,10,100]$ (green), $[200,10,200]$ (orange), and $[500,50,500]$ (black). A portion of the classical result for the largest coupling strength, $[500,50,500]$, is also shown for comparison (black dashed line). For the largest coupling, the decay is fitted to a power law, $C_v (t) \sim t^{-d}$. The power law exponents are: $d_{\rm qm} = 0.86$, $d_{\rm cl} = 0.91$ for $|9,10,16\rangle$, $d_{\rm qm} = 0.93$, $d_{\rm cl} = 0.92$ for $|20,7,9\rangle$ and $d_{\rm qm} = 0.91$, $d_{\rm cl} = 0.93$ for $|25,4,9\rangle$. (D) Fragmentation of the ZOBS $|9,10,16\rangle$ (top row), $|20,7,9\rangle$ (middle row) and $|25,4,9\rangle$ (bottom row) with  coupling strengths increasing  from left to right. Here, $\Delta E_i = E_i - E_{\bf{v}}^{(0)}$, where $E_i$ is the eigenenergy of the $i^{\rm th}$-state and $E_{\bf{v}}^{(0)}$ is energy of the ZOBS. See the text for details.}
\label{fgr:smoothed_surv_prob}
\end{figure}

In Fig.~\ref{fgr:states_location} three example ZOBS are shown  - $|9,10,16\rangle$ located near the junction formed by the ${\bf r}_{2}$ and ${\bf r}_{3}$ resonances, $|25,4,9\rangle$ located in the vicinity of the ${\bf r}_{1}-{\bf r}_{3}$ junction, and $|20,7,9\rangle$ which is located away from the junctions. The choice of ZOBS near the  different junctions has to do with the fact that the two junctions are not dynamically equivalent. Figure~\ref{fgr:smoothed_surv_prob} summarizes  the quantum and classical IVR dynamics in terms of the time-smoothed survival probability
\begin{equation}
    C_{\bf v}(t) \equiv \frac{1}{t} \int_{0}^{t} P_{\bf v}(t') dt'
    \label{smoothpoft}
\end{equation}
associated with the three example ZOBS. Comparing the  behaviour of eq.~\ref{smoothpoft} for the three example states brings out certain crucial differences in the IVR dynamics. Firstly, the ZOBS located in the vicinity of the junctions exhibit decays even at the smallest coupling strengths when compared to the ZOBS located away from the junctions. This is due to the larger number of near-degenerate states available at the junctions. In these regimes the ZOBS are involved in a coherent resonance-assisted tunneling process involving the near-degenerate states. A clear confirmation comes from the fact that the analogous classical $C_{\bf v}(t)$ for $|25,4,9\rangle$ does not decay. Interestingly, the  classical $C_{\bf v}(t)$ for $|9,10,16\rangle$ does decay for the smallest coupling strengths, albeit on a much faster time scale. This, as evident from Fig.~\ref{fgr:web_df_nloc}(a), is due to the substantial chaos  near the ${\bf r}_{2}-{\bf r}_{3}$ junction as compared to the ${\bf r}_{1}-{\bf r}_{3}$ junction. Thus, for small coupling strengths the classical and quantum IVR pathways are very different. Secondly, the effect of the junctions is highlighted by the non-monotonic decay of $C_{\bf v}(t)$ in Fig.~\ref{fgr:smoothed_surv_prob}A and C even at intermediate coupling strengths. Notice that in comparison to the junction ZOBS, for the ZOBS $|20,7,9\rangle$ the extent of non-monotonicity is much reduced. The local increase in $C_{\bf v}(t)$ is clearly due to the recurrences in the survival probability. For a ZOBS near junction the stronger recurrences come from dynamical tunneling involving the ZOBS and the ``dark" zeroth-order states located in the vicinity of the junction. For sufficiently large values of the coupling strengths one can see from Fig.~\ref{fgr:smoothed_surv_prob}A and C that the recurrences are suppressed and the classical and quantum $C_{\bf v}(t)$ exhibit similar power law scalings. However, it is interesting that the long time limit of $C_{\bf v}(t)$, essentially the dilution factor $\sigma_{\bf v}$ in the absence of exact degeneracies, is larger for the ZOBS near junctions. It appears, although largely conjectural at this point, that the combination of dynamical tunneling and the classical Nekhoroshev stability results in the quantum IVR dynamics of such junction ZOBS to be much more localized in comparison to the classical IVR dynamics. 

In Fig.~\ref{fgr:smoothed_surv_prob}D the corresponding local density of states (LDOS) or ``spectrum" $\rho(E) = \sum_{\alpha} |C_{\alpha {\bf v}}|^{2} \delta(E-E_{\alpha})$ are also shown   with increasing coupling strengths. As expected, the LDOS is nearly Gaussian for the largest coupling strength considered here. However, for small to intermediate couplings the typical off-resonant superexchange pathways\cite{stuchebrukhovmarcus1,stuchebrukhovmarcus2,SrihariPRE2005} are apparent. As argued previously, such vibrational superexchange pathways precisely correspond to dynamical tunneling occurring between near-degenerate zeroth-order states on the Arnold web\cite{SrihariPRE2005}. Note that the LDOS also indicates that dynamical tunneling (both resonance and chaos assisted) mechanism at the ${\bf r}_{1}-{\bf r}_{3}$ junction persist longer than in the case of the ${\bf r}_{2}-{\bf r}_{3}$ junction. Clearly, the IVR dynamics for states near different junctions are different, and at present there is little known about any a priori criteria  leading to the identification of specific junctions that may play an important role in the quantum IVR dynamics.  

\begin{figure*}[tbp]
\centering
  \includegraphics[width=1\linewidth]{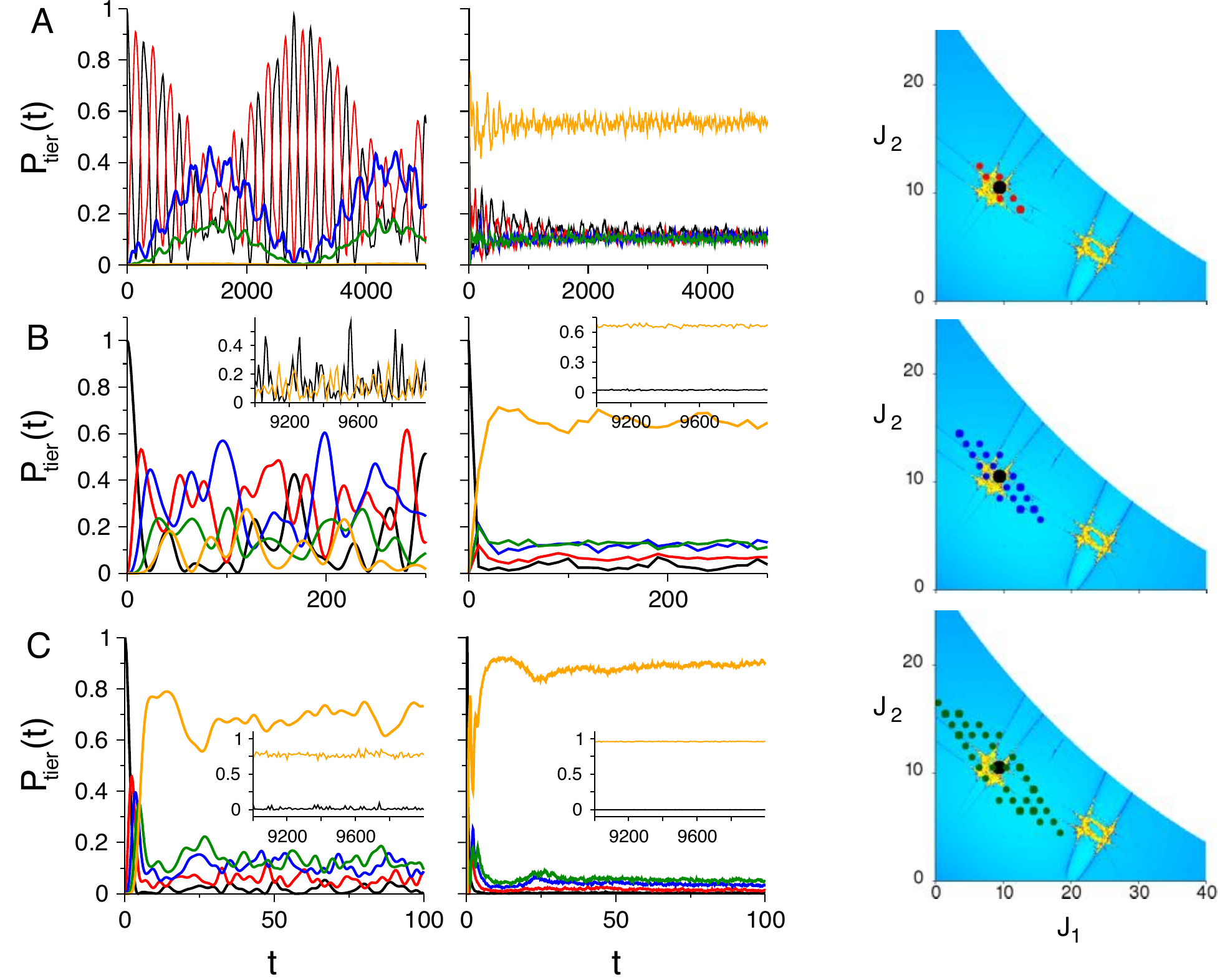}
  \caption{Quantum (left column) and classical (middle column) probability flow to different tiers from the ZOBS $|9,10,16\rangle$ for different coupling strengths, (A) $[20,1,20]$ (B) $[100,10,100]$  (C) $[500,50,500]$. Note that the survival probability of the ZOBS is shown in black while the total probability of the tier one ($6$ states), two ($18$ states) and, three $3$ ($38$ states) are shown in red, blue and, green respectively. The probability corresponding to the rest of the ``bath" states (beyond tier three) are shown in orange. Insets in B and C show the survival and bath probabilities at long times. In the right column, the location of the projection of the various tier states are shown on the Arnold web for the choice of angle slice $(\theta_1,\theta_2,\theta_3)=(\pi/2,\pi/2,\pi/2)$. Note that the various tier states have differing $v_{3}$ values and the webs do not reflect the varied coupling strengths considered in the $P_{\rm tier}(t)$ plots. See  text for discussions.}
  \label{fgr:tier_flow}
\end{figure*}

A more explicit demonstration of the competition between the classical and quantum IVR pathways, and the critical role of the resonance junctions, is presented in Fig.~\ref{fgr:tier_flow}. Here, as an example, the IVR out of the ZOBS $|9,10,16\rangle$  is shown in terms of the flow of energy into the states belonging to different tiers. The criteria in eq.~\ref{pertind} is used to sort the different tiers; for simplicity we keep the first three tiers and group the rest of the states as ``bath". For low coupling strengths one can clearly observe in Fig.~\ref{fgr:tier_flow}A an almost coherent energy flow back and forth between the ZOBS and the first tier states. Energy flows into the second and third tier states on  significantly longer timescales. However, note that the IVR into second and third tier states is strongly correlated and not sequential. These coherent oscillations in the tier probabilities are signatures of the resonance-assisted tunneling mediated IVR pathways near the junction\cite{KarmakarKesh2018}. The corresponding classical tier flows are completely different from the quantum case. As argued above, the set of dark states participating dominantly via dynamical tunneling in the quantum IVR  are inaccessible to the classical IVR dynamics. However, the fast classical IVR observed in Fig.~\ref{fgr:tier_flow}A represents a local ``thermalization" due to the chaotic dynamics near the ${\bf r}_{2}-{\bf r}_{3}$ junction (cf. Fig.~\ref{fgr:web_df_nloc}(a)). Note that while the quantum energy flow is almost exclusively restricted to the first three tiers, classically substantial energy flow does occur into the bath states. In addition, it is clear from Fig.~\ref{fgr:tier_flow}A that ``backflow" into tiers (cf. Fig.~\ref{tiersketch} indicated by loops) occurs to a considerable extent in such cases. This is a consequence of the trapped dynamics near the junction which is further accentuated due to dynamical tunneling. However, surprisingly, this significant inter-tier correlation persists even for moderately large couplings, as shown in Fig.~\ref{fgr:tier_flow}B. Evidently, even at this larger coupling strength, the stabilizing influence of the junction continues to play an important role in the IVR dynamics and questions the appropriateness of thinking in terms of an irreversible relaxation of energy through a hierarchical set of tier states. It is only at the largest coupling strength shown in Fig.~\ref{fgr:tier_flow}C that the dynamical influence of the junction appears to weaken and a reasonable correspondence between the classical and quantum tier probabilities, including the bath probabilities, can be observed. An important observation here is that Fig.~\ref{fgr:tier_flow}C corresponds to most of the QNS having $N_{\rm loc} \geq 1$ (cf. Fig.~\ref{fgr:web_df_nloc}(c)) and the model, as indicated by the dilution factor distribution in Fig.~\ref{fgr:df_hist}, moving closer to the QET regime. 

\subsection{Resonance junctions: seeds of nonstatisticality?}

The notion that regions in phase space where there is a confluence of several independent resonances can lead to the phenomenon of resonance stabilization is not new, and has been known since the famous work\cite{Nekhoroshev1977} of Nekhoroshev nearly half a century ago. However, the dynamical consequences of such stabilization over physically relevant timescales in various systems is only now beginning to be appreciated\cite{PsakauskasPRL08,manikandankesh14,pankajkesh2015,SKPKYKS2020}. As aptly stated by Lochak\cite{Lochak1993}, coming to terms with the notion of local stability in the vicinity of the resonance junctions requires overcoming two strong prejudices. The first prejudice, arising from ``linear thinking" about resonances, has to do with associating resonances with a loss of stability, which can only be restored by dissipative forces. In fact, finite time stability can manifest in Hamiltonian systems with $f \geq 3$, without the need to invoke non-conservative forces. The second prejudice comes from the KAM theorem itself which advertises the eternal stability of only the very non-resonant regions in the phase space. Actually, highly resonant regions are also stable, but only for long times and not eternally\cite{pankajkesh2015,SKPKYKS2020}. Interestingly, trapping near resonance junctions can also be seen in the early work\cite{milczewskietal96} of Milczewski, Diercksen, and Uzer where the Arnold web for the hydrogen atom in crossed electric and magnetic fields is computed. Clearly, what constitutes ``long time" vis-\`{a}-vis the physically relevant timescales is important and needs to be sorted out for specific systems. Nevertheless, for molecular systems one typically has $f \geq 3$ and, given that the junctions are dense on the Arnold web, there is a good chance that the IVR dynamics of  experimentally prepared bright states are bound to be influenced by a certain set of junctions. 

The results and discussions in the context of the simple $f=3$ effective model Hamiltonian of eq.~\ref{ham_qm_ccm} clearly put the spotlight on the resonance junctions. In fact, several earlier studies on IVR dynamics on different molecular systems seem to hint at the role of the resonance junctions. Examples include the beautiful and very detailed classical dynamical studies on overtone induced dissociation of H$_{2}$O$_{2}$ by Uzer, Reinhardt, and Hynes\cite{UzerReindardtH2O2CPL,UzerReindardtH2O2JCP} and the isomerization dynamics of HCN by Hutchinson and coworkers\cite{HolmeHutchinsonHCN1985}. In both the examples, with the former being an exceptionally brave attempt given that it is a $f=6$ system, deep insights into the dynamics of energy flow came from adopting the classical nonlinear resonance perspective. Could it be that certain ``surprising" aspects in these older studies can be understood better now in terms of the Arnold web structure? We certainly think so, but further technical progress and conceptual advances are needed before taking on such systems. An example, however,  which is worth mentioning and which is set up well for analysis along the lines presented in this perspective is that of IVR in the highly excited carbonyl sulfide (OCS) molecule. The pioneering study by Martens, Davis, and Ezra\cite{Martensetal1987} of the IVR dynamics near the dissociation threshold of the planar OCS  already provided examples for trapping near resonance junctions\footnote[5]{As an aside we mention that many of the novel ideas and suggestions regarding transport and bottlenecks that were proposed in this work\cite{Martensetal1987} have yet to be explored in detail,  particularly the quantum manifestations.}. A more recent  incisive  study\cite{PsakauskasPRL08,Pasketal2009} by Pa{\v{s}}kauskas, Chandre, and Uzer on the same system established that the trajectories get temporarily trapped around families of two-dimensional invariant tori organized around specific periodic orbits. It is relevant to note here that periodic orbits are expected at the resonance junctions. However, further work is required to establish whether this recent observation is related to specific resonance junctions. Certainly, based on the earlier studies by Vela-Arevalo and Wiggins\cite{arevaloPhD}, there are good reasons to believe so. Note that the work\cite{PsakauskasPRL08,Pasketal2009} of Pa{\v{s}}kauskas, Chandre, and Uzer also provides a possible link between the NHIM and the escape mechanism associated with  such significantly trapped trajectories which can result in non-RRKM behaviour. The recent detailed quantum energy flow dynamics study\cite{PerezArceOCS} in OCS by P\'{e}rez and Arce confirms the non-RRKM dynamics even at energies near the dissociation threshold. Interestingly, apart from connecting to the ideas of ETH and MBL, P\'{e}rez and Arce argue that the non-RRKM behaviour arises despite the molecule being fairly close to thermalizing. Based on our recent work\cite{pankajkesh2015,SKPKYKS2020}, such an observation points to the role of trapping regions in the phase space - possibly even related to the influence of the resonance junctions. 

Although, as mentioned above,  OCS  has served as a paradigmatic molecule for understanding various aspects of the IVR dynamics, a clear quantum-classical correspondence at the level of detail shown in Figs.~\ref{fgr:web_df_nloc}-\ref{fgr:tier_flow} is still lacking. 
Recent studies\cite{manikandankesh14,KarmakarKesh2018}, however, are starting to focus on establishing a more explicit correspondence between the features on the Arnold web and the quantum IVR dynamics. For example, motivated by the extensive experimental studies by Gruebele and coworkers\cite{LeeGruebele2006,ChowdaryGruebele2007,ChowdaryGruebele2009} and Rashev and Moule\cite{rachevmoule2008}, analysis of an effective spectroscopic Hamiltonian for the thiophosgene (SCCl$_{2}$) molecule reveals that the junctions can lead to fairly nontrivial IVR dynamics. Thus, edge (overtone) states in the QNS can undergo greater extent of IVR as compared to a near-degenerate interior (combination) state in the QNS. Moreover, removing a specific anharmonic resonance can lead to enhanced IVR - a counterintuitive result that can be rationalized based on the local trapping dynamics near a specific junction\cite{manikandankesh14}.
The work\cite{manikandankesh14} exploring the IVR dynamics in a spectrally congested region of SCCl$_{2}$ and the more recent work\cite{SKPKYKS2020} on the  unimolecular dissociation of model triatomic systems (the Bunker models) are starting to provide compelling evidence for the central role that such resonance junctions play in determining the onset of statistical behaviour. 


\begin{figure}
\centering
  \includegraphics[width=1\linewidth]{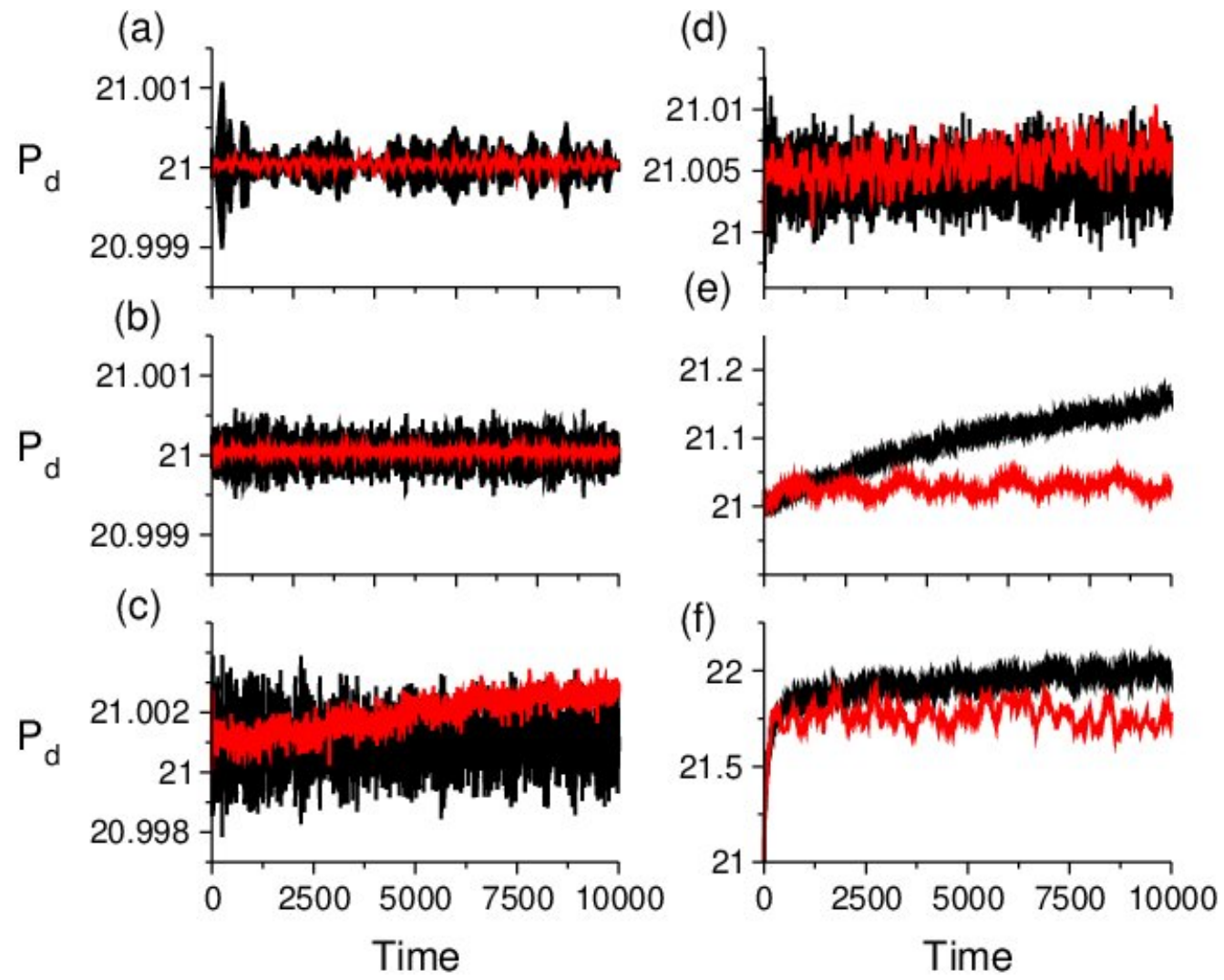}
  \caption{Approximate conserved classical (black) and the corresponding quantum (red) polyad, $P_d$ defined in eq.~\ref{polyr1r3}, as a function of time for the initial state $|25,4,9\rangle$. The different resonance coupling strengths are, (a) $[5,1,5]$, (b) $[20,1,20]$, (c) $[50,5,50]$, (d) $[100,10,100]$, (e) $[200,10,200]$, (f) $[500,50,500]$. Note the different y-axis scales.}
  \label{fgr:polyad_25_4_9}
\end{figure}

It is, of course, crucial to ask if there are any spectroscopic signatures of the resonance junctions. One potential candidate is the existence of approximately conserved quantities. Trapping in the vicinity of a specific junction implies that the dynamics is mainly influenced by the independent resonances that intersect to form the junction. Thus, near the resonance junction, using an appropriate canonical transformation one can construct a local constant of the motion.  For example, for the model Hamiltonian eq.~\ref{ham_qm_ccm}, one expects the approximate conservation of the quantity 
\begin{equation}
P_{d} \equiv (J_1 + J_3)/2 + J_2 
\label{polyr1r3}
\end{equation}
near the ${\bf r}_{1}-{\bf r}_{3}$ junction. The corresponding quantum object $(v_{1} + v_{3})/2 + v_{2}$ is nothing but a polyad. Note, as pointed out earlier, in the absence of the ${\bf r}_{2}$ resonance, $P_{d}$ is exactly conserved. In Fig.~\ref{fgr:polyad_25_4_9} we explicitly show the existence of such an approximate polyad during the time evolution of the ZOBS $|25,4,9\rangle$ located close to the junction.  For low values of the resonance coupling strengths, the trapping near the junction leads to the polyad being well conserved. However, even  for the larger coupling strengths, but the two junctions still distinct, the dynamics (both classical and quantum) shows the polyad to be approximately conserved. Comparison of Figs.~\ref{fgr:polyad_25_4_9}(c),(d) with the Arnold webs shown in Fig.~\ref{fgr:web_df_nloc}(a) is particularly instructive. For the larger coupling strength case in Fig.~\ref{fgr:polyad_25_4_9}(e), the quantum dynamics seems to obey the polyad conservation much more than  classical dynamics. This again points to the quantum dynamics being more localized near the junctions, presumably due to the persistence of resonance-assisted tunneling\cite{brodieretalprl01,eltschkaetalprl05,locketal10} as evident from the LDOS shown in Fig.~\ref{fgr:smoothed_surv_prob}D, when compared with the classical localization. As expected, for the largest coupling strength considered here and shown in Fig.~\ref{fgr:polyad_25_4_9}(f), both the quantum and classical polyads show non-conservation of $P_{d}$ at early times. This is consistent with Fig.~\ref{fgr:web_df_nloc}(a) showing the large scale overlap of the regions associated with the two junctions. However, interestingly, Fig.~\ref{fgr:polyad_25_4_9}(f) shows the quantum polyad fluctuating around a different value over fairly long times. The dynamical significance of this observation, and whether it is a general feature at other resonance junctions, is currently not known.

\begin{figure*}
\centering
  \includegraphics[width=0.7\linewidth]{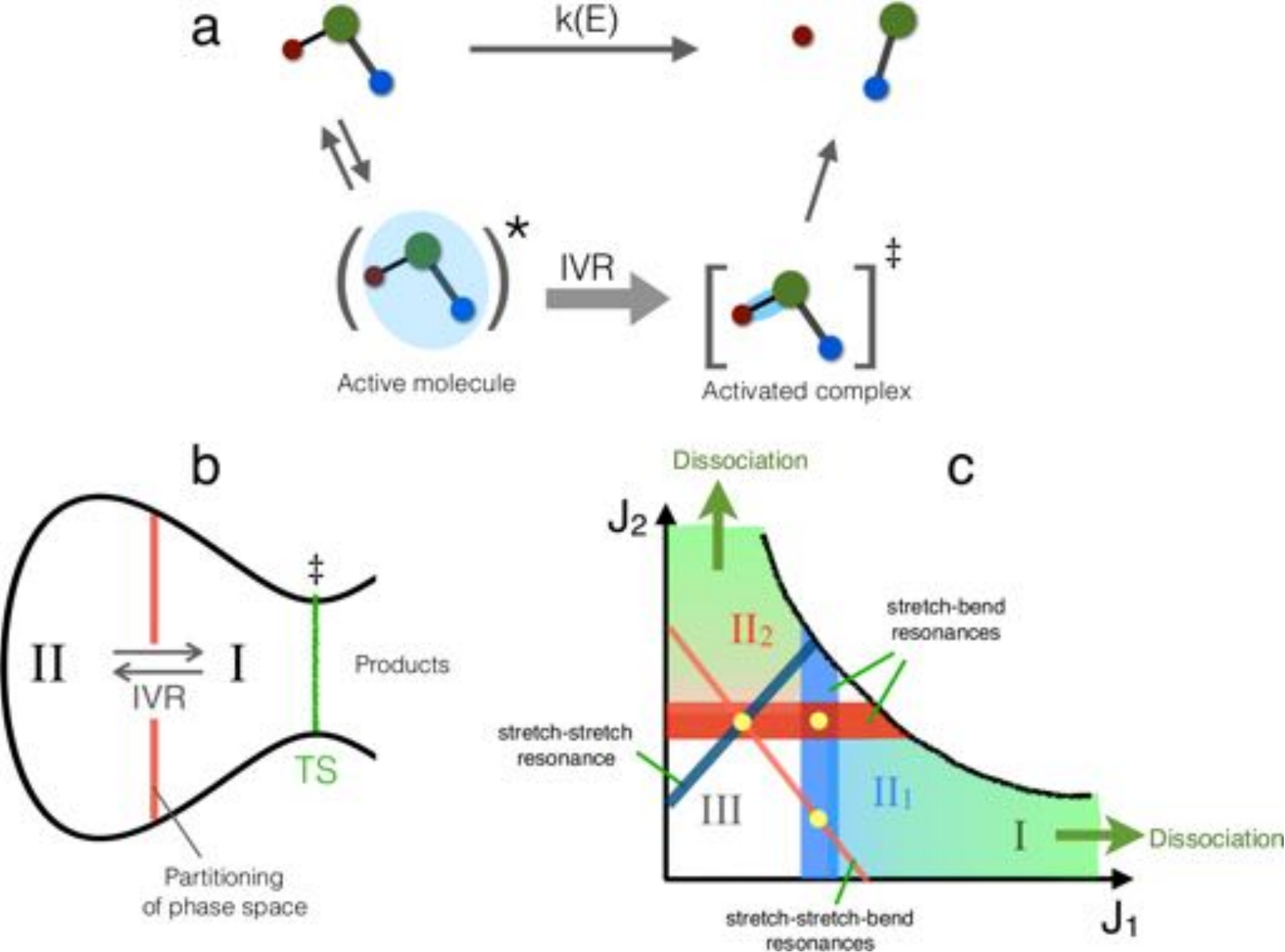}
  \caption{Schematic illustration of the various viewpoints on unimolecular reactions. (a) The standard Lindemann-Hinshelwood-Marcus mechanism for the dissociation of a triatomic molecule. (b) A kinetic scheme to account for nonstatistical dynamics based on a partitioning of the phase space into regions with distinct dynamical behaviours. (c) The state (action) space representation showing the constant energy surface decorated by the nonlinear resonances involving the stretching and the bending modes. The various independent resonances can intersect to form junctions (yellow circles).}
  \label{fgr:unirate_sketch}
\end{figure*}

A similar analysis, not shown here, near the ${\bf r}_{2}-{\bf r}_{3}$ junction reveals the approximate conservation of the polyad $P_{d}' \equiv 4 v_{1}/3 + 2 v_{2} + v_{3}$. Generalization to multiplicity$-r$ resonance junction for $f > 3$ cases can similarly be made. Given the importance of polyads in interpreting the spectral features, one anticipates that significant trapping near a junction should result in observing patterns in the overtone spectra. However, as opposed to the usual single resonance polyad patterns, the junction  based dynamical polyad patterns cannot be discerned by inspecting the harmonic mode frequencies alone. In this context note that the observation of  nested IVR timescales in the dynamics of highly excited acetylene (C$_{2}$H$_{2}$) in an early work\cite{HolmeLevine1989} by Holme and Levine seems to be a signature of  trapping near resonance junctions. If this is indeed correct then the connection made by Holme and Levine between their observed time-scale separations and the experimental spectral features at different resolutions may be an experimental manifestation of Nekhoroshev stability! Furthermore, nearly three decades ago, Engel and Levine\cite{EngelLevine1989} have already made an attempt to correlate the IVR dynamics of ZOBS prepared by stimulated emission pumping (SEP) with their proximity to specific resonance junctions on the Arnold web. Perhaps a re-look at the molecules  investigated by Levine and coworkers\cite{EngelLevine1989,HolmeLevine1989} is in order.

\section{IVR and unimolecular dissociation reactions}
\label{RRKMornot}

The preceding discussion on the role of resonance junctions, and other features on the Arnold web, to the IVR process has a direct bearing on unimolecular rate processes. In particular, the standard Lindemann-Hinshelwood-Marcus kinetics (cf. Fig.~\ref{fgr:unirate_sketch}(a) for a schematic) assumes that the IVR, corresponding to a timescale $\tau_{\rm IVR} \equiv k_{\rm IVR}^{-1}$,  that takes the active molecule to the activated complex (transition state) is sufficiently fast\footnote[6]{Note that in this perspective we are not concerned with the first step in unimolecular kinetics involving collisional activation of the reacting molecule. This is still an active area of research, particularly when one does not or cannot invoke the strong collision approximation. For example look at the recent feature article by Jasper\cite{AhrenJasperJPCA}.} when compared to the timescale $\tau_{R} \equiv \nu_{R}^{-1}$ associated with crossing of the transition state by the activated complex. Clearly, the results of the previous sections suggest that trapping near the resonance junctions can delay the dynamics sufficiently to violate the crucial RRKM assumption. In other words, the IVR does not repopulate the TS fast enough to maintain a quasi steady state population near the TS. As argued by Leitner\cite{leitnerACP2005}, such delays in the IVR process, even for situations wherein the QET lies below the bottleneck energy, can lead to rates significantly smaller than the RRKM prediction. Thus, the RRKM rate is modified\cite{LeitnerIJQC1999,Leitner2015}  as $k(E) = \kappa(E) k_{\rm RRKM}(E)$ with
\begin{equation}
    \kappa(E) = \frac{k_{\rm IVR}(E)+\omega_{\rm coll}}{k_{\rm RRKM}(E) + \omega_{\rm coll} + \nu_{R}(E)}
\end{equation}
where we have denoted the typical collision frequency, within a strong collision approximation, by $\omega_{\rm coll}$. One now has to explicitly calculate $k_{\rm IVR}(E)$ - a nontrivial dynamical task. Note that LRMT does provide a way to compute the energy flow rates as a function of the total energy. 

It is important to mention that within the above viewpoint  it is quite possible to have an exponential decay of the survival probability of an initial microcanonical ensemble $N(0)$
\begin{equation}
    S(t) \equiv \frac{N(t)}{N(0)} = e^{-k(E)t}
\end{equation}
but with $k(E) \neq k_{\rm RRKM}(E)$. In the above $N(t)$ is the number of initial conditions that remain undissociated at time $t$. On a related note, recent studies\cite{jayeehase2019} by Hase and coworkers  also brings out the crucial point that ``fitting an experimental thermal unimolecular rate constant versus temperature and pressure with the Hinshelwood-Lindemann-RRKM model does not unambiguously identify the unimolecular reactant as a RRKM molecule". In fact, very early on Marcus, Hase, and Swamy\cite{MarcusHaseSwamy} proposed a kinetic model to account for deviations from single exponential kinetics. This model imagines the reactant phase space to be partitioned (cf. a schematic in Fig.~\ref{fgr:unirate_sketch}(b)) into two or more regions with finite IVR rates between them. The partitions, implicitly associated with various anharmonic resonances, are then associated with multiexponential behaviour of the survival
\begin{equation}
    S(t) = \sum_{j} f_{j} e^{-k_{j}(E)t}
    \label{survprob}
\end{equation}
and the related, more fundamental, quantity called as the lifetime distribution
\begin{equation}
    P(t) \equiv \frac{dS(t)}{dt} = \sum_{j}f_{j} k_{j}(E) e^{-k_{j}(E) t}
    \label{lifetimeprob}
\end{equation}
However, Ezra, Waalkens, and Wiggins\cite{ezrawaalkenswigginsHCN2009} in their detailed and precise analysis of the $f=3$ HCN isomerization reaction have clearly highlighted the subtleties involved. Using the gap time distribution perspective due to Slater\cite{Slaterbook,SlaterJCP1956} and Thiele\cite{ThieleJCP1962,ThieleJCP1963}, combined with the recently advanced NHIM-based phase space TST, they show that
\begin{enumerate}
\item Both power law behaviour at intermediate times and an exponential  decay at times much longer than the mean gap time are seen in the integrated gap time distribution, which is closely related to the lifetime distribution in eq.~\ref{lifetimeprob} above. 
\item There exist trapped regions in the reactive phase space and accounting for them, based on a ``simple-minded" relative phase space density argument, leads to significant over correction to the usual RRKM rate.
\end{enumerate}
A couple of comments on the observations above are pertinent at this stage. Already an earlier work\cite{Shojiguchi2008} by Shojiguchi, Li, Komatsuzaki, and Toda on a model $f=3$ system suggests that the  first point above has links to the dynamics on the Arnold web. Although the precise nature of dynamical barriers were not identified, it was clear from their analysis that the intermediate time power law and long time exponential behaviour of the survival probability arose from trajectories transitioning from certain regions in the Arnold web to the vicinity of resonance junctions. Secondly, the issue of over correcting the RRKM estimate is also related to the features on the Arnold web in the reactant well, since reactive trajectories that cross the dividing surface, but nevertheless get temporarily trapped near the junctions, lead to non-ergodicity. Specifically, correlated motion near the junctions invalidate  the assumption that phase space densities can be partitioned into sub-densities corresponding to dynamically distinct behaviours. Similar concerns, related to the stickiness and vague tori in $f=2$ systems, have been noted by De Leon and Berne\cite{deleonbernejcp81} and others\cite{ShirtsReinhardt1982,Haseetal1984}. 

From the above brief discussion one can conclude, as also emphasized in an earlier study\cite{Shojiguchi2008} Shojiguchi et al, that for a clearer understanding of the rate processes it is imperative to investigate the nature of the Arnold web in the reactant phase space regions. More importantly, relating the topology of the stable and unstable manifolds of the NHIM in the bottleneck regions, that control the entry into and exit from the reactant phase space region, with the nature of the Arnold web in the reactant region should provide a sophisticated understanding of the regimes of validity of statistical rate theories. An equally important task is to provide unambiguous connections between the dynamical barriers on the Arnold web and the unimolecular decay observables. Undertaking such a task requires an appropriate model $f=3$ system which, apart from being accessible to the modern tools of nonlinear dynamical systems theory, embodies the key aspects of unimolecular reactions. Examples such as HCN isomerization\cite{HolmeHutchinsonHCN1985,WaalkBurbWigJCP2004,LiTodaTamikiJCP2005,GongMaRiceJCP2005,WaalkBurbWigPRL2005,ezrawaalkenswigginsHCN2009}, OCS dissociation\cite{CarterBrumerJCP1982,Martensetal1987,arevaloPhD,shchekinova04,PsakauskasPRL08,Pasketal2009,PerezArceOCS}, and several others mentioned above are indeed potential candidates. However, in our opinion, the set of model triatomic molecules studied\cite{Bunker1,Bunker2} by Bunker nearly half a century ago provide a good starting point for several reasons. Firstly, the Bunker models provide a fairly comprehensive set to test the various aspects and assertions regarding the connection between IVR and the Arnold web structures. Secondly, detailed quantum dynamical study of the models to ascertain the extent of classical-quantum correspondence are yet to be made. Finally, these models provide an ideal platform to address the, still outstanding, issue of determining the necessary and sufficient condition for a molecule to be in the RRKM regime. Combined with the quantum dynamical studies, the last point may lead to a clear connection with the QET.  

\subsection{The Bunker models: phase space perspective}

In a series of tour de force computational studies\cite{Bunker1,Bunker2} Bunker considered the Hamiltonian
\begin{equation}
H({\bf q},{\bf p}) = \sum_{i=1}^{3} \left[\frac{1}{2} G_{ii}({\bf q}) p_{i}^{2} + V_{i}(q_{i}) \right] + \sum_{i < j=1}^{3} G_{ij}({\bf q}) p_{i} p_{j}
\label{Bunker_ham}
\end{equation}
to investigate the classical unimolecular dissociation dynamics of the various model non-rotating triatomic molecules in terms of the lifetime distributions. The  coordinate dependent G-matrix elements in eq.~\ref{Bunker_ham} are defined as follows:
\begin{eqnarray}
G_{11}({\bf q}) &=& \frac{1}{\mu_1} \nonumber \\
G_{22}({\bf q}) &=& \frac{1}{\mu_2} \nonumber \\
G_{33}({\bf q}) &=& \frac{1}{\mu_{1}q_{1}^{2}} + \frac{1}{\mu_{2} q_{2}^{2}} - \frac{\cos q_{3}}{M q_{1} q_{2}} \\
G_{12}({\bf q}) &=& \frac{\cos q_{3}}{M} \nonumber \\
G_{13}({\bf q}) &=& - \frac{\sin q_{3}}{M q_{2}} \nonumber \\
G_{23}({\bf q}) &=& - \frac{\sin q_{3}}{M q_{1}} \nonumber
\end{eqnarray}
with $\mu_{k} \equiv 2m_{k}M/(m_{k} + M)$ being the relevant reduced masses.
Bunker considered various harmonic and anharmonic models for the potentials $V_{i}(q_{i})$ and classified them as being RRKM or not (cf. Table~\ref{tbl:bunker}).
Based on his computations, Bunker clearly brought out the role of anharmonicity, provided insights into the relation between IVR and the random gap time assumption, and suggested the possibility of multiexponential lifetime distributions. Furthermore, he concluded that models with large disparity in the masses and frequencies were prone to being non-RRKM and that for most models IVR was complete in about $\sim 10$ ps. 

Given that Bunker's computations were entirely classical, and on a seemingly simple $f=3$ system, an interesting question is whether one can a priori predict the models that would exhibit non-RRKM behaviour. In other words, to what extent can one rationalize the results of Bunker  based on ideas from nonlinear dynamics? Perhaps more ambitiously\footnote[7]{Don Bunker himself would have been appalled, perhaps, by this deductionism! For example, in his accounts article\cite{BunkerACR1974} he says ``But the workers in this field will be able to stave off deductionism. The reason is that, inductively, compliance with the basic assumption, random re-distribution of energy, is not obligatory but must be reconfirmed for every molecule." }, the aim would be to predict the models in Table~\ref{tbl:bunker} that exhibit RRKM behaviour without performing detailed dynamical calculations. The answer came about fifteen years later in the form of a beautiful analysis by Oxtoby and Rice\cite{OxtobyRice} of the nonlinear resonances in the models of Bunker.  For reasons mentioned below the model considered by Oxtoby and Rice\cite{OxtobyRice} 
\begin{equation}
H({\bf q},{\bf p}) = \sum_{i=1}^{3} \left[\frac{1}{2} G^{(0)}_{ii} p_{i}^{2} + V_{i}(q_{i}) \right] + \epsilon \sum_{i < j=1}^{3} G^{(0)}_{ij} p_{i} p_{j}
\label{Bunkeroxrice_ham}
\end{equation}
with $\epsilon=1$ is inspired by that of Bunker, except that the coordinate dependent $G$-matrix elements are replaced  with their equilibrium values.  
The bond stretching modes ($i=1,2$) were modeled by Morse oscillators
\begin{equation}
V_{i}(q_{i}) = D_{i} [1-\exp(-\alpha_{i}(q_{i}-q_{i}^{0}))]^{2}
\end{equation}
and the bending mode ($i = 3$) was  considered to be harmonic 
\begin{equation}
V_{3}(q_{3}) = \frac{1}{2G_{33}^{(0)}}\omega_{3}^{2} (q_{3}-q_{3}^{0})^{2}
\end{equation}
The various Hamiltonian parameters are chosen consistent with the values given in Table~\ref{tbl:bunker}. Thus, given the harmonic frequencies $\omega_{K}$ and the dissociation energies $D_{k}$ of the stretching modes in Table~\ref{tbl:bunker}, the Morse parameter is fixed as $\alpha_{k} = \omega_{k}/\sqrt{2 D_{k} G_{kk}^{(0)}}$.

\begin{table*}[htbp]
\small
  \caption{\ Parameters for the different Bunker models\cite{Bunker2}}
  \label{tbl:bunker} 
  \begin{threeparttable}

  \begin{tabular*}{1\textwidth}{@{\extracolsep{\fill}}cccccc}
    \hline \vspace{1mm}

    Model & Masses (amu)\tnote{*} & Energies (kcal mol$^{-1}$) & Equilibrium geometry\tnote{**} & Frequencies (cm$^{-1}$) & RRKM or not\\ \vspace{1mm}
         & $[m_1, M, m_2]$ & $[D_1, D_2]$ & $[q_1^0, q_2^0,q_3^0]$ & $[\omega_1, \omega_2, \omega_3]$ & \\

    \hline \vspace{1mm}

$2$       & $[14,14,16]$ & $[114,84.7]$  & $[1.15,1.23,\pi]$     & $[2228,1266,591]$    & \cmark \\
\vspace{1mm}
$3$       & $[14,14,16]$ & $[114,60]$    & $[1.15,1.23,\pi]$     & $[2228,1266,591]$    & \cmark \\
\vspace{1mm}
$6$       & $[16,16,16]$ & $[24,24]$     & $[1.278,1.278,2.039]$ & $[1112,1040,632]$    & \cmark \\
\vspace{1mm}
$7$       & $[24,16,24]$ & $[60,60]$     & $[1.2,1.2,\pi]$       & $[1505,752.5,752.5]$ & \cmark \\
\vspace{1mm}
$10$      & $[1,24,24]$  & $[114,84.7]$  & $[1.15,1.23,\pi]$     & $[5086,1377,1028]$   & \xmark \\
\vspace{1mm}
$8$       & $[1,24,24]$  & $[114,60]$    & $[1.15,1.23,\pi]$     & $[5086,1377,1028]$   & \xmark \\
\vspace{1mm}
$8 \rm A$ & $[1,12,12]$  & $[114,60]$    & $[1.05,1.55,\pi]$     & $[3500,1505,1110]$   & \xmark \\
\vspace{1mm}
$8 \rm B$ & $[1,12,12]$  & $[114,60]$    & $[1.05,1.55,\pi]$     & $[1750,1505,1110]$   & \xmark \\
\vspace{1mm}
$8 \rm C$ & $[12,12,12]$ & $[114,60]$    & $[1.05,1.55,\pi]$     & $[3500,1505,1110]$   & \xmark \\
\vspace{1mm}
$8 \rm D$ & $[1,12,12]$  & $[60,114]$    & $[1.05,1.55,\pi]$     & $[3500,1505,1110]$   & \xmark \\
    
   \hline
\end{tabular*}
\begin{tablenotes}\footnotesize
\item[*] The central mass $M$ is connected to the masses $m_{1}$ and $m_{2}$ with equilibrium bond lengths $q_{1}^{0}$ and $q_{2}^{0}$ respectively.
\item[**] $q_1^0$ and $q_2^0$ are in angstroms and $q_3^0$ is in radians.
\end{tablenotes}

\end{threeparttable}

\end{table*}


The assumption $G_{ij}({\bf q}) \approx G_{ij}^{(0)}$ allows for transforming eq.~\ref{Bunkeroxrice_ham} to the action-angle representation 
$H({\bf J},{\bm \theta}) = H_{0}({\bf J}) + \epsilon V({\bf J},{\bm \theta})$
with the zeroth-order part given by 
\begin{equation}
H_{0}({\bf J}) = \sum_{k=1,2} \omega_{k} J_{k} \left(1 - \frac{\omega_{k}}{4D_{k}} J_{k} \right) + \omega_{3} J_{3}
\label{zerothH}
\end{equation}
  The modes are coupled by the perturbation
\begin{eqnarray} 
V({\bf J},{\bm \theta}) &= &\sum_{l,m = 1}^{\infty} f^{(12)}_{lm}({\bf J}) \left[\cos(l \theta_{1}-m \theta_{2}) - \cos(l \theta_{1}+m \theta_{2})\right]   \nonumber\\
                                   &+& \sum_{l=1}^{\infty} g^{(13)}_{l}({\bf J}) \left[\sin(l \theta_{1}-\theta_{3})+\sin(l \theta_{1}+\theta_{3})\right]  \nonumber \\
                                   &+& \sum_{m=1}^{\infty} g^{(23)}_{m}({\bf J}) \left[\sin(m \theta_{2}-\theta_{3})+\sin(m\theta_{2}+\theta_{3})\right] \label{resterms}
\end{eqnarray}
with the Fourier coefficients $f_{lm}({\bf J})$ and $g_{l}({\bf J})$ being functions of the parameters of the Hamiltonian in eq.~\ref{Bunkeroxrice_ham}. For a detailed description of the transformation and the expressions for the various Fourier coefficients we refer the readers to the recent work\cite{SKPKYKS2020}.
Thus, eq.~\ref{Bunkeroxrice_ham} can be transformed into the form of eq.~\ref{ham_actangrep}, making the model amenable to the various tools and techniques of nonlinear dynamics. Neither the original Bunker model in eq.~\ref{Bunker_ham} nor the model Hamiltonians for OCS\cite{CarterBrumerJCP1982}, HCN\cite{HolmeHutchinsonHCN1985}, or H$_{2}$O$_{2}$\cite{UzerReindardtH2O2JCP}  can be transformed easily into the action-angle representation. Although the assumption can be questioned, the insights gained from analyzing eq.~\ref{Bunkeroxrice_ham} far outweigh the slight differences in the various dynamically computed quantities. In fact, the classification of models as being RRKM or not appears to remain intact despite the assumption.  Note that as opposed to the model effective Hamiltonian in eq.~\ref{ham_qm_ccm}, the above Hamiltonian has an infinity of nonlinear resonances. A further simplification was made by Oxtoby and Rice\cite{OxtobyRice} by assuming that the bending degree of freedom can be neglected, thus the terms $g_{l}^{(13)} = 0 = g_{m}^{(23)}$ in eq.~\ref{resterms}, resulting in a $f=2$ system involving only the anharmonic stretching modes. At this stage, 
motivated by the simple criteria of Chirikov that overlapping nonlinear resonances generate widespread chaos, Oxtoby and Rice\cite{OxtobyRice} surmised that the Bunker models which exhibit RRKM behaviour must be in the Chirikov regime. The reasoning essentially had to do with possible links between chaos and ergodicity on the constant energy surface.  Hence, they concluded that the bottlenecks to IVR  ``arise due to the existence of isolated, non-overlapping, nonlinear resonances which can trap the energy of the molecule for a large number of vibrational periods." In this context, the penetrating classical-quantum correspondence study\cite{raikay1984} of the dissociation dynamics of collinear ($f=2$) triatomics by Rai and Kay is worth noting. 

Despite the fact that the analysis of Oxtoby and Rice represents the closest we have got to an a priori determination of whether a given molecule is in the RRKM regime or not, the fact remains that the notion of ``isolated resonance" is not tenable in the original $f=3$ Bunker model. Indeed, as sketched in Fig.~\ref{fgr:webgeom} the various resonances in eq.~\ref{resterms} will decorate the constant energy surface to form an intricate Arnold web in the reactant region.  Note that Oxtoby and Rice themselves were well aware of this and state in their paper\cite{OxtobyRice} ``For two interacting oscillators (the case we have discussed so far), the zeroth order energy surface is one dimensional. This case is simple in that there is only one path from a given point to any other, so that narrow resonances do not overlap. For $n >2$ oscillators, however, the energy hypersurface is of dimension $n - 1$, and centers of resonances are not points but $n - 2$ dimensional surfaces. For a given energy, one can study the positions and extent of overlap of the resonances on the energy hypersurface; the degree to which the energy hypersurface is covered by overlapping resonances determines the extent to which the system may be described stochastically. However, such a many-dimensional picture rapidly becomes very complicated even for quite small molecules, so the question arises of whether it can be replaced by an approximate one-dimensional picture". 

So, will the analysis of the $f=3$ system by including the bending degree of freedom confirm the expectations of Oxtoby and Rice? A recent study\cite{SKPKYKS2020} of Bunker's model $6$ provides the answer by correlating the features on the Arnold web with the dissociation dynamics.  We refer the reader to the original article\cite{SKPKYKS2020} for details and provide a summary of the significant findings. Firstly, the application of a Chirikov type criterion is fraught with difficulties. However, numerically, the Arnold webs provide the kind of intuitive picture advanced by Oxtoby and Rice. Secondly, the bending degree of freedom does play a crucial role via the involvement of a dominant resonance junction formed by the two independent $1$:$1$ stretch-bend nonlinear resonances. More significantly, the temporary trapping near the junction correlates directly with the long time tails seen in the lifetime distributions. Thirdly, the influence of the junction persists beyond the extensive resonance overlap regime. Hence, it seems plausible that the dynamical trapping near the junctions could be one key source for non-RRKM behaviour. It would be interesting to subject the Bunker model to a LRMT analysis and see if the QET corresponds to a weakening influence of the junction.

As a final point, it is interesting that the results of the Bunker model $6$ indicate that the trapped dynamics near the junctions is strongly correlated\cite{SKPKYKS2020}. From a molecular viewpoint, at such junctions the IVR process is rather fast and involves all three modes.  Therefore, the junctions are ``local pockets" of  facile IVR, seemingly what is precisely needed to ensure statisticality. However, the resonance stabilization at the junction ensures that the lifetime distribution, and hence the gap time distribution, will be non-exponential. This is seemingly at odds with the conclusion made by Uzer, Hynes, and Reinhardt\cite{UzerReindardtH2O2JCP} that ``lack of a sufficiently perfect resonance path to the reaction coordinate augurs well for RRKM behaviour". 

\subsection{Bunker's classification - an FLI perspective}
\label{Bunkerbelowdiss}

In sec.~\ref{three_resmodel} we demonstrated a close  correspondence between the features on the Arnold web and the various measures that signal QET for the model effective Hamiltonian. A rather detailed study of Bunker's model $6$, as summarized above, provides insights into the features on the Arnold web that control the onset of statisticality. How general then is the association of the trapping near junctions with non-RRKM behaviour? Clearly, a thorough classical and quantum dynamical studies of the rest of the Bunker models shown in Table~\ref{tbl:bunker} is warranted. However, as a preliminary result, we show that the classification  of the various models by Bunker as RRKM or not is reflected in the FLI-based Arnold web features.

\begin{figure}[t]
\centering
 \includegraphics[width=1\linewidth]{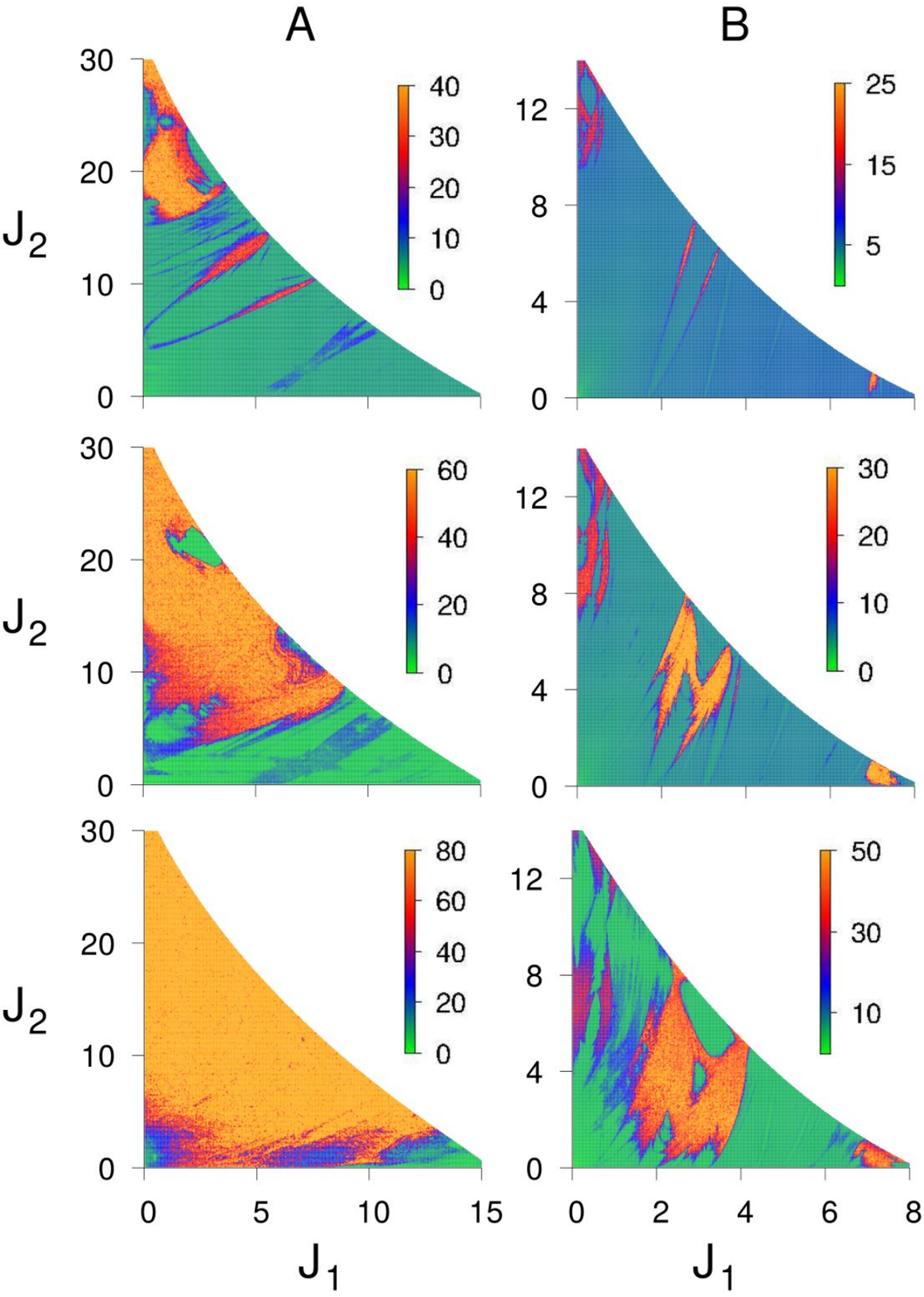}
 \caption{Evolution of the Arnold web as a function of $\epsilon$ at energy, $E = 0.9D_2$ for Bunker's model $2$ (column A) and model $8D$ (column B).  Top, middle, and bottom panels correspond to $\epsilon = 0.1, 0.25$, and $0.5$ respectively. The FLI values shown are obtained by integrating trajectories to the final time $t=40$. The webs shown correspond to the choice of angle slice $(\theta_1,\theta_2,\theta_3)=(\pi/2,\pi/2,0)$.}
 \label{fgr:onset_of_chaos}
\end{figure}

Before showing the result we note that the equilibrium $G$-matrix assumption is retained in the following. Consequently, for certain models in Table~\ref{tbl:bunker}, the equilibrium value of the bend angle $q_3^{(0)} = \pi$ results in a $f=2$ system, since $G_{13}^{(0)} = G_{23}^{(0)} = 0$. Here, we rectify this in an ad hoc fashion by fixing $q_3^{(0)} = 2$ radians for such models. Another approach, as adopted by Oxtoby and Rice\cite{OxtobyRice}, is to consider the next non vanishing term in the expansion of the $G({\bf q})$ about the equilibrium geometry. A little bit of thought shows that our ad hoc inclusion of the third degree of freedom will nevertheless be dynamically similar to the full Bunker model, given that we are at energies far away from the equilibrium geometry. Thus, we do not claim that the models represent any specific molecule. The emphasis here is to qualitatively illustrate the connection between dynamics on the web and the ``RRKMness" of the models.  

First, we consider the two models, $8D$ and $2$, predicted by Bunker to be non-RRKM and RRKM respectively. In Fig.~\ref{fgr:onset_of_chaos}A and B the Arnold webs for model $2$ and $8D$ respectively are shown as a function of the coupling strength (cf. eq.~\ref{Bunkeroxrice_ham}) $\epsilon$. Note that the webs are shown at a total energy below the dissociation energy. Specifically, the energies in Fig.~\ref{fgr:onset_of_chaos} correspond to $E = 0.9 D$ with $D$ being the lower of dissociation energies indicated in Table~\ref{tbl:bunker}. As a function of $\epsilon$, a much more rapid onset of chaotic dynamics for the model $2$  is evident from Fig.~\ref{fgr:onset_of_chaos}A. In contrast, Fig.~\ref{fgr:onset_of_chaos}B clearly shows that there are significant structures that persist even at moderate coupling strengths for model $8D$.  Estimates based on the zeroth-order Hamiltonian suggest that the for $\epsilon \sim 0.5$, the lower FLI value region around $(J_{1},J_{2}) \approx (3,6)$  corresponds to a junction. 

\begin{figure}[t]
\centering
  \includegraphics[width=1\linewidth]{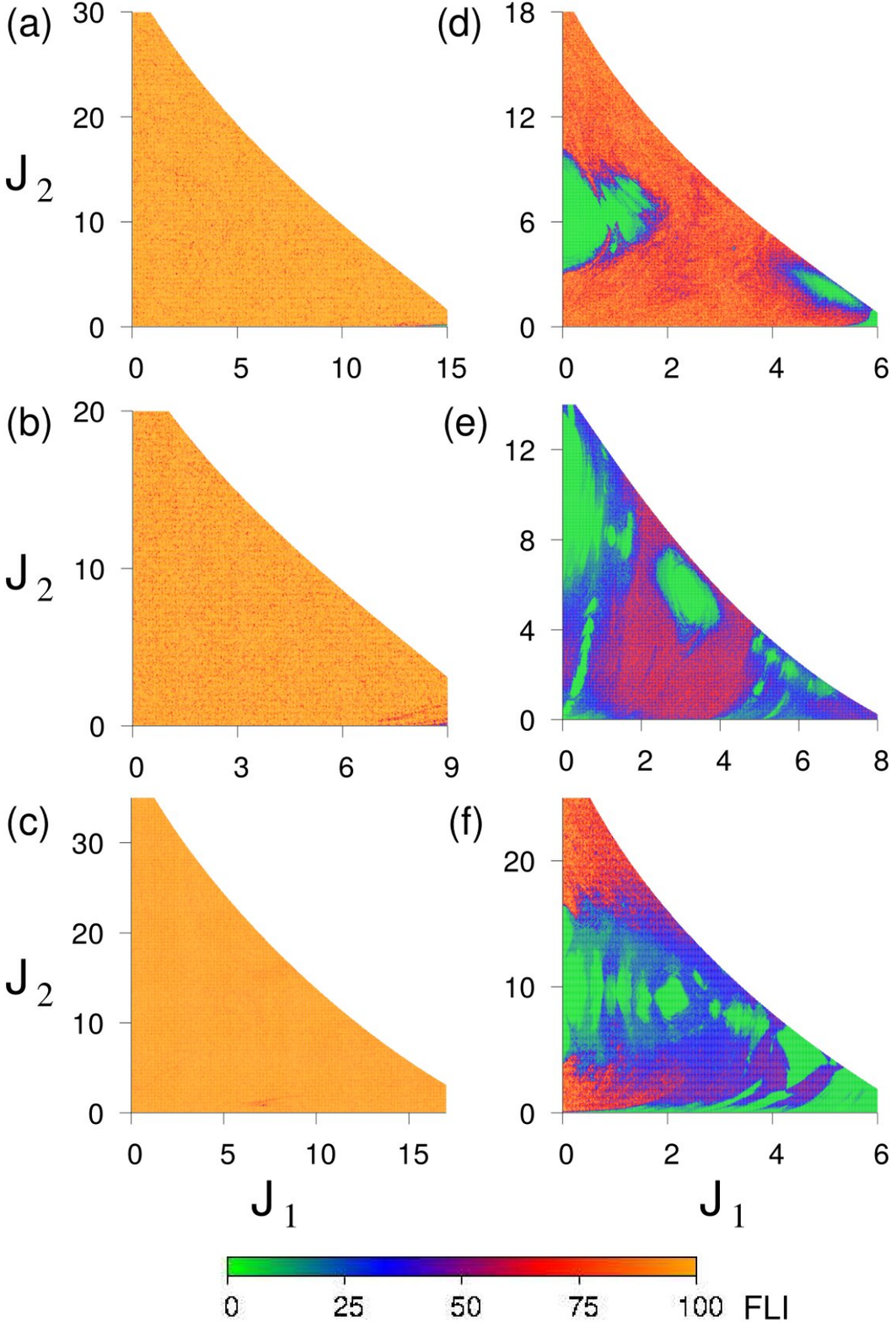}
  \caption{The Arnold webs for $\epsilon=1$ projected on $(J_1,J_2)$ space at $90$\% of the dissociation energy for Bunker models (a) $2$, (b) $3$, (c) $7$, (d) $8A$, (e) $8D$, and (f) $10$. Final time of integration and angle slice as in Fig.~\ref{fgr:onset_of_chaos}.}
  \label{fgr:bunker_web}
\end{figure}

In Fig.~\ref{fgr:bunker_web} the webs for six different Bunker models are shown for the full coupling strength i.e, $\epsilon = 1$. As can be seen from Fig.~\ref{fgr:bunker_web}(a), model $2$ shows almost no structure and extensive chaos. On the other hand Fig.~\ref{fgr:bunker_web}(e) for model $8D$ indicates that the junction continues to influence the IVR dynamics. Of the four other Bunker models shown in Fig.~\ref{fgr:bunker_web}, models $3$ and $7$ are in the RRKM regime according to Bunker's computations. Consistently, one can see that these models already show strongly chaotic dynamics with no evidence of junctions being present. The other two models $8A$ and $10$ are predicted to be non-RRKM by Bunker and clearly, as shown in Fig.~\ref{fgr:bunker_web}(d) and (f) respectively, residual structures in the phase space can be seen.

\begin{figure}[t]
\centering
  \includegraphics[width=0.7\linewidth]{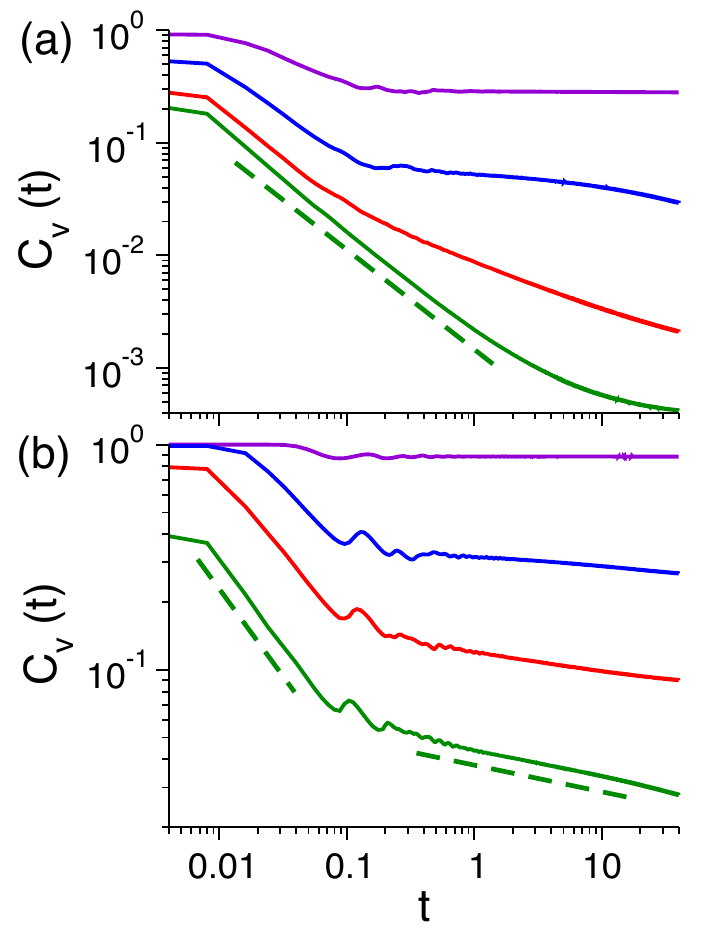}
  \caption{Smoothed classical survival probabilities for initial states (a) $|J_1,J_2,J_3\rangle = |6,9,4\rangle$ (model $2$) and (b)  $|3,4,3\rangle$ (model $8D$) for  couplings strengths $\epsilon = 0.1$ (violet), $0.25$ (blue), $0.5$ (red), and $1.0$ (green). Power law fitting for the $\epsilon = 1$ case are shown as green dashed lines.}
  \label{fgr:bunker_sp}
\end{figure}

As mentioned above, analyzing the Bunker models in terms of the scaling theory\cite{schofieldwolynes1993} of Schofield and Wolynes is of some interest. As the quantum dynamics of the models  are yet to be studied in detail, in Fig.~\ref{fgr:bunker_sp} the classical smoothed survival probabilities $C_{v}(t)$ (cf. eq.~\ref{smoothpoft}) for initial states corresponding to $|v_1,v_2,v_3 \rangle = |6,9,4 \rangle$ (for model $2$) and $|3,4,3 \rangle$ (for model $8D$) are shown for increasing coupling strengths. Based on the understanding gained from the $C_{v}(t)$ results for the model effective Hamiltonian  shown in Fig.~\ref{fgr:smoothed_surv_prob}, we expect the presence of a junction to lead to multiple time scales.  This expectation is borne out in Fig.~\ref{fgr:bunker_sp}(b) for model $8D$. The early time decay scales as $C_v(t) \sim t^{-0.76}$ and at longer time as $C_v(t) \sim t^{-0.12}$.  The origin of such multiple exponents, as shown in the previous study on the SCCl$_2$ system\cite{manikandankesh14}, is due to the heterogeneous nature of the phase space as seen explicitly in Fig.~\ref{fgr:onset_of_chaos}-\ref{fgr:bunker_web}. In contrast, for model $2$ the result in Fig.~\ref{fgr:bunker_sp}(a) shows a transition to single exponent decay $C_v(t) \sim t^{-0.88}$. Note that the exponent is close to being consistent with the expected $t^{-(f-1)/2}$ diffusive scaling, suggesting an almost complete transition to the statistical regime. 

\section{Concluding thoughts and future challenges}

Our results and discussions up until now have sought to bring out the explicit classical-quantum correspondence involving the dynamics of energy flow in polyatomic molecules.  Decades ago, Crim\cite{Crim90} pointed out that ``The notion of exciting a particular motion in a molecule in order to control the breaking of a chemical bond is very appealing, but its implementation requires a sophisticated understanding of vibrationally energized molecules and unimolecular reactions". What we have attempted to establish in this Perspective is  the sort of ``sophisticated understanding" that may be required to fully understand the mechanism and appreciate the complexity of IVR.  For sure we now have powerful tools to understand the classical dynamics in large dimensional phase spaces, and therefore extend/question the beautiful insights into IVR and reaction dynamics obtained in lower dimensional systems.  However, there are still challenges that require conceptual advances and, in this context, connections to other areas of research allows for a broader perspective on the relevant issues. We end this Perspective by alluding  to the outstanding issues and the interesting connections.

\subsection{Connections}

\subsubsection{Dynamical astronomy} Several of the results shown here and more recent studies indicate that simply dissecting a dynamical system into a regular part and a chaotic part is perhaps too simplistic. Not all chaotic trajectories are same in systems with mixed regular-chaotic phase spaces, and in the $f \geq 3$ case various possibilities exist. For a fully chaotic trajectory, the total energy is the only conserved quantity. However, for a partially chaotic trajectory, apart from the energy, one or more conserved quantity may exist. There is a debate\cite{Froeschle71,LLbook} on the veracity of such claims. Nevertheless, in a recent study Muzzio\cite{Muzzio2017} shows the presence of partially chaotic dynamics near a resonance junction. Interestingly, much of the discussions and debate in this context have been in the field of dynamical astronomy. Thus, the notions of stable chaos\cite{MilaniNobili1992,Milani1997}, partially chaotic orbits\cite{Contopoulos78,Muzzio2017}, and stickiness in dynamical astronomy should be relevant in the molecular context as well\footnote[8]{Indeed, there is an interesting parallel between unimolecular dissociation reactions and the discussions in dynamical astronomy involving the connections\cite{morbfroes96} between the Lyapunov and escape times of asteroids.}. Note that this issue has been recognized fairly early in the context of  IVR. Examples include the studies by  Kosloff and Rice\cite{KosloffRice1981}, Hamilton and Brumer\cite{HamiltonBrumer1985}, and Kuz'min et al\cite{Kuzminetal1986}. More recent studies\cite{lange2016} by Lange, B\"{a}cker, and Ketzmerick is starting to shed light on the distinctly different mechanism of power law trapping in $f=3$ systems. Consequently, a better understanding of the vibrational predissociation dynamics of weakly bound van der Waals clusters, modeled as open $4$D maps by Gaspard and Rice\cite{gaspardrice89} and Gillilan and Ezra\cite{GillilanEzra1991} , may be possible. Interestingly, the recent work\cite{dasbacker20} of Das and B\"{a}cker on the $3$D ABC-map shows that power law behaviour of the classical Poincar\'{e} recurrence probability can be approximately decomposed into a sum of dominant exponential contributions due to different nonlinear resonances on the Arnold web - could this observation lead to a dynamical interpretation of the weights of the exponentials in eq.~\ref{survprob}, and hence connecting the scaling and the kinetic model approaches to IVR? Presumably, a more detailed analysis with the Nekhoroshev theorem as a platform needs to be done for a precise answer.

\subsubsection{Thermalization of isolated quantum systems} Earlier in this perspective we have hinted at the connections of statistical IVR to the issues of ETH and MBL. In fact, given the similarity of the effective spectroscopic Hamiltonians and the tight-binding or Bose-Hubbard models, we expect several ideas and measures being developed in the two areas to have significant intersections. This is due to the reason that the scaling theory\cite{schofieldwolynes1993} and LRMT descriptions of IVR\cite{schofieldwolynes1994,leitnerwolynes1997} arose from observing the analogy\cite{loganwolynes1990} to the phenomenon of Anderson localization. Moreover, as discussed by Leitner\cite{LeitnerEntropy2018} recently, the quantum localization in the QNS is an example of MBL. For instance, the recently proposed criteria for MBL by Serbyn, Papi\'{c}, and Abanin\cite{SerbynPapicAbaninPRX} in terms of the measure ${\cal G}(\epsilon, L)$ is quite similar to the $N_{\rm loc}$ measure invoked in the LRMT and BSTR models for IVR. Similarly, the existence of approximately conserved quantities in the vicinity of the junctions may have connections to the notion of local integrals of motion (LIOMs)\cite{SerbynPapicAbanin2013,HuseNandOganesyan2014}. In fact, Nekhoroshev type estimates have been used recently in the context of understanding the quantum quench dynamics of 1D Bose gases\cite{BrandinoCauxKonikPRX}.  A point to note here is that in most of the MBL studies it is natural, and advantageous, to utilize the disorder averaging technique. For IVR in isolated molecules there is no disorder to average over. Nevertheless, there have been recent studies\cite{papic2015,MoessnerPRL17,nieuwenrafaelpnas19,SchulzPollPRL19,KunoNJP2020} on the possibility of observing MBL in disorder free systems. 

\subsection{Controlling IVR from the Arnold web viewpoint}

As mentioned in the Introduction, much of the early control ideas emerged from an implicit assumption about the typical IVR timescales being too fast to perform mode-specific chemistry.  However, the effective IVR dimension of the QNS is small and, typically, an intermediate time power law behaviour is observed\cite{GruebelePNAS1998}. Gruebele\cite{GruebeleTCA2003} has argued that these aspects can be used to freeze the IVR over chemically significant timescales. A different viewpoint comes from an early work\cite{OhtsukiJCP2001} by Ohtsuki et al wherein it was shown that optimal control fields can utilize the IVR dynamics to efficiently transfer population from an initial to a target final quantum state. It was shown that the dark states, populated by a $2$:$1$ stretch-bend IVR, play a crucial role in the control process. Interestingly, Ohtsuki et al visualized the control dynamics in the QNS to gain insights into the control mechanisms. 

More recently, Shi and Schlegel\cite{ShiSchlegel2019} have analyzed the dissociation dynamics of ClHCO$^{+}$ in the presence of two intense laser pulses. They used a wavelet-based time-frequency analysis of the Born-Oppenheimer molecular trajectories to gain detailed insights into the mechanism of the various dissociation channels. The wavelet analysis yields direct information on the IVR occurring during the fragmentation process. Although not discussed in this Perspective, the wavelet approach has been utilized to understand the IVR mechanism in a number of systems\cite{arevalowiggins,ChandreWiggUzerPhysD,Bach2006,manikandan2009,manikandankesh14,pankajkesh2015,SKPKYKS2020}. We refer the readers to earlier reviews\cite{arevalowiggins,ChandreWiggUzerPhysD} for details on the power and utility of the technique. In fact, a rather detailed analysis of the dissociation dynamics of HCN\cite{sethi2012} in terms of the Arnold web structure, determined using wavelet transform, reveals a wealth of information.  It was shown that the frequency ratio space $(\Omega_{\rm CH}/\omega_{F},\Omega_{\rm CN}/\omega_{F})$ with $\omega_{F}$ being the field frequency clearly brings out the mechanism of IVR in the presence of the external field. Dissociation dynamics was shown to be regulated by a transition state like region and sticky regions corresponding to the ``decoupling" of the field from the two stretching modes. More recent work\cite{LopezBenitoBorondojcp2016} by Borondo and coworkers presents further detailed analysis and clearly brings out the correlation between the dissociation probabilities and the features on the Arnold web. 

It should not come as a surprise that the Arnold web perspective can be utilized to control the reaction dynamics. After all, the web encodes the ``IVR traffic" of the molecule and  external fields that couple to the vibrations can direct the traffic. In particular, junctions with their associated stable chaos provide ``waypoints" to redirect the IVR. As a fairly simple example, consider a ZOBS $|v_{1},v_{2}\rangle$ which classically is in a rather broad resonance zone $\alpha_{1} \Omega_{1} \approx \alpha_{2} \Omega_{2}$. In the absence of an external field the across resonance diffusion may lead to dissociation. However, one can now use an external ac-field of frequency $\omega_{F}$ to ensure that the ZOBS is located in the vicinity of the junction formed by the resonances $m \Omega_{1} \approx f_{1} \omega_{F}$ and $n \Omega_{2} \approx f_{2} \omega_{F}$, provided the condition $m f_{2} \alpha_{2} \approx n f_{1} \alpha_{1}$ is satisfied. If so, then the ZOBS can be stabilized near the junction.

\subsection{The grand challenge: going beyond three degrees of freedom}

We started this Perspective by citing examples of nonstatistical dynamics in the context of reactive intermediates, thermalization or lack thereof in the presence of solvents, and other systems with $f \gg 3$. Additionally, a recent work\cite{ChenetalNatChem20} by Chen et al establishes that nonstatistical dynamics can manifest in mechanochemical reactions.  This clearly has consequences for techniques that utilize  external forces to sample multidimensional potential energy surfaces\cite{Maedaetal2015} and free energy landscapes\cite{Pauletal2019}. In this context, and given that we have mostly restricted our attention\footnote[9]{By no means are we suggesting that analysis of $f=3$ system is straightforward. A case in point is the role of nonstatisticality in the mass-independent fractionation of ozone. As emphasized recently by Babikov and coworkers\cite{BabikovetalJCP18}, a fully ab initio quantum explanation for the various dynamical effects are far from easy. In fact, the dynamical origins of the so called $\zeta$-effect and $\eta$-effect\cite{GaoMarcusScience01,Marcus2013} are only now beginning to be understood\cite{BabikovetalJCP18}.}
to isolated systems with $f=3$, several questions arise. Is the analysis based on Arnold webs even relevant or useful for systems with $ f \gg 3$? How would one construct and visualize such webs? Can the qualitative insights be translated into quantitative corrections for the various rate constants?

Answering the various questions above is far from easy. Perhaps, similar concerns were raised a few decades ago in terms of generalizing  our understanding of classical-quantum correspondence in IVR dynamics from $f=2$ to $f=3$ case. An ``easy", and potentially evasive, answer is that the  number of modes that actually are relevant to the IVR dynamics is fairly small even in very large molecules. Consequently, effective Hamiltonians obtained from a CVPT analysis or from fitting of experimental spectra are sufficiently low dimensional that they can be analyzed using the techniques presented here. Having said that, below we give brief, and necessarily preliminary, thoughts on the questions.

The answer to the first question above is clearly in the affirmative. For example, Shudo, Ichiki, and Saito\cite{shudosaito2006} have argued that the slow relaxation in liquid water dynamics may be linked to the presence of bottlenecks in the large dimensional phase space. The very recent study\cite{dasgreen19} by Das and Green demonstrates, using finite time Lyapunov exponents,  the slowing down of chaos near liquid-vapor critical point for non-associated liquids.  A second example comes from the analysis of the Fermi-Pasta-Ulam-Tsingou (FPUT) model by Pettini and Landolfi\cite{PettiniLandolfi1990}. They conclude that, depending on the initial condition, even for $f \gg 3$ one can have very long-lived metastable states. It is relevant to note at this juncture that the role of resonances to the equilibration timescale in the FPUT model has recently been highlighted by Onorato and coworkers\cite{OnoratoPNAS2015,BustamanteOnorato2019}  and the quantum aspects have been studied by Burin et al\cite{Burinetal2019}. 

A potential strategy for addressing the second issue is to exploit the time scale separation between high and low frequency modes to construct ``adiabatic" Arnold webs. Inspired by the recent vibrational conical intersection picture of IVR by Hamm and Stock\cite{HammStock2012}, it may be possible to construct an approximate framework based on the adiabatic Arnold webs for understanding thermalization in large systems. In this context it is interesting to note that Tesar et al\cite{tesaretal13} have utilized a Marcus electron transfer theory approach to IVR in large polyatomics. Connections can also be made to the stochastic pump model for diffusion in this context\cite{leitwolynesprl97,Leitwolynescpl97,manikandanKS07,Malyshev2010}.

The last question raised above is crucial from the control perspective as well, particularly in systems with post-TS bifurcations\cite{EssHouk2008,haretantillo2017,Tantillo2018}. The reactive island theory\cite{deleonRI1,deleonRI2,deleonRI3} of De Leon and coworkers and the approximate phase space bottlenecks approach\cite{davisgray1986,grayrice1987} of Gray, Rice, and Davis have already shown in $f=2$ systems that quantitative corrections to rates can certainly be made. Thus, combined with the tremendous advances being made in identifying the phase space TS from the NHIM perspective, quantitative characterization of the transport on the Arnold web in the reactant regions for $f \geq 3$ is necessary. Apart from the  wavelet-based time-frequency analysis\cite{arevalowiggins,ChandreWiggUzerPhysD} and a bevy of sophisticated chaos indicators\cite{Froeschle1997,Froeschle2002,Barrio05,Barrio06,Cincotta03,Skokos01,Skokos07,Funketal04,gottwald04,gottwald09,Skokos2016}, recent progress towards visualizing the multidimensional phase spaces\cite{katsanikas11,katsanikas13,richter2014,lange2014,firmbach18} and the potential for detecting the manifolds and web structure using measures such as the Lagrangian descriptor\cite{mendozamancho2010,manchowiggins2013,lopesino2017} certainly seem promising for taking the powerful classical-quantum correspondence technique to the next level.

\section*{Acknowledgements}

S. Ke. benefited greatly from discussions with David Leitner, Mikito Toda, Tetsuya Taketsugu, Akira Shudo, Arnd B\"{a}cker, and Martin Gruebele.  He also gratefully acknowledges the hospitality of the Faculty of Science, Hokkaido University, Max-Planck Institute for Physics of Complex Systems, Dresden, and the School of Mathematics, University of Bristol where parts of this Perspective were written. This work is funded  by the Science and Engineering Research Board of India (EMR/2016/0062456). S. Ka. thanks the University Grants Commission (UGC), India for a doctoral fellowship. We acknowledge the IIT
Kanpur High Performance Computing center for providing computing time.

\section{Appendix: Fast Lyapunov indicator}
\label{FTLI}

In Hamiltonian systems, the coexistence of both the regular and chaotic regions in the phase space makes the dynamics very rich and complicated. The characterization of the nature (regular or chaotic) of any trajectory (or region) is therefore essential to predict stability. The visualization of the phase space is a direct way to identify different regions. For one degree of freedom (DOF) system, the phase space being two dimensional is easy to visualize. For two DOF system, the phase space can be visualized using the two dimensional Poincar\'{e} surface of section (PSOS) technique. However for three and higher DOF system, the PSOS itself becomes $\geq 4$ dimensional and the visualization of the phase space becomes difficult. 

The characterization of trajectories using dynamical indicators is a possible approach. In the context of dynamical systems, the Lyapunov exponents (LE)\cite{Skokos2016}  as indicator have a long history. The Lyapunov exponent is defined as
\begin{equation}
 \lambda = \lim_{t\to\infty, ||\Delta x(0)||\to\ 0} \Big(\frac{1}{t}\Big) \ln \frac{||\Delta x(t)||}{||\Delta x(0)||}
\end{equation}
where $||\Delta x(t)||$ is the distance between two nearby trajectories. By definition the LE is positive for chaotic and zero for regular trajectories. However, computation of LE requires a sufficiently long time and hence computationally challenging for higher DOF systems. Again, the accurate long time numerical integration is also computationally very demanding. Consequently, several finite time chaos indicators\cite{Skokos2016} have been introduced over the past few decades. In this Perspective we use the fast Lyapunov indicator (FLI) approach\cite{Froeschle1997,FroeschleScience2000} for constructing the various Arnold webs. The FLI measures the norm of the tangent vector associated with a trajectory during the time evolution with the aim of providing a fast distinction between dynamically different trajectories.

For a dynamical system 
\begin{equation}
    \frac{d\mathbf{X}}{dt} = \mathbf{F}(\mathbf{X})
\end{equation}
with $\mathbf{X} = \{x_1, x_2, \ldots, x_i, \ldots\}$ being the dynamical variables, the FLI\cite{Froeschle1997} is defined as 
\begin{equation}\label{def1}
    \mathrm{FLI}(\mathbf{X}(0), \mathbf{v}(0), t) = \mathrm{log}||\mathbf{v}(t)||
\end{equation}
where, $\mathbf{v}$ is a tangent vector. The evolution of the tangent vector is obtained from the variational equation 
\begin{equation}
    \frac{d\mathbf{v}}{dt} = \Big(\frac{\partial \mathbf{F}}{\partial \mathbf{X}} \Big) \mathbf{v}
\end{equation}
However, for technical reasons the alternative definition\cite{Lega2001}
\begin{equation}\label{def2}
    \mathrm{FLI}(\mathbf{X}(0), \mathbf{v}(0), t) = \sup_{0 < t^{'} <t} \mathrm{log}||\mathbf{v}(t^{'})||
\end{equation}
is usually preferred, and we also use eq.~\ref{def2} to compute the FLI.  The variation of FLI with time for regular and chaotic trajectories is distinctly different and hence different types of dynamical behaviour can be identified, usually within timescales $t_f$ much shorter compared to that of required for computing the LEs. For bounded systems  a fixed, appropriately chosen time $t_f$ is usually sufficient to discriminate between the different types of dynamics. However, note that for unbounded or open systems trajectories can escape or dissociate over a wide range of time scales. Therefore, a single fixed  total integration time $t_f$ might be insufficient and further analysis is required\cite{SKPKYKS2020}. 

For an integrable system (also for a KAM trajectory), the FLI increases as $\log(t)$. For a chaotic trajectory, as expected, FLI increases linearly with $t$. An advantage of the FLI approach is that one can also distinguish between a regular nonresonant trajectory (KAM trajectory) and a regular resonant trajectory\cite{Froeschle2002}. Note that, an arbitrary choice of the initial tangent vector can lead to spurious structures\cite{Barrio09} on the web. To avoid this, we follow the prescription by Barrio et al\cite{Barrio09} and choose the initial tangent vector as $-\nabla H/||\nabla H||$.

We now illustrate the FLI method  for the H\'enon-Heiles system\cite{HenonHeiles}. The Hamiltonian is of the form
\begin{equation}\label{HH_Ham}
    H = \frac{1}{2}(p_x^2 + p_y^2 + x^2 + y^2) + x^2y - \frac{1}{3}y^3
\end{equation}
The Hamiltonian equations of motion can be expressed as
\begin{gather}
 \frac{d}{dt}\begin{bmatrix} x \\ y \\ p_x \\ p_y \end{bmatrix}
 =
  \begin{bmatrix}
   p_x \\
   p_y \\
   - x - 2xy \\
   - y - x^2 + y^2
   \end{bmatrix}
\end{gather}
and the relevant variational equations are
\begin{gather}
 \frac{d}{dt}\begin{bmatrix} v_1 \\ v_2 \\ v_3 \\ v_4 \end{bmatrix}
 =
  \begin{bmatrix}
   0 & 0 & 1 & 0 \\
   0 & 0 & 0 & 1 \\
   - 1 - 2y & -2x & 0 & 0 \\
   -2x & - 1 + 2y & 0 & 0
   \end{bmatrix}
   \begin{bmatrix} v_1 \\ v_2 \\ v_3 \\ v_4 \end{bmatrix}
\end{gather}
\begin{figure}[t]
\centering
  \includegraphics[width=1\linewidth]{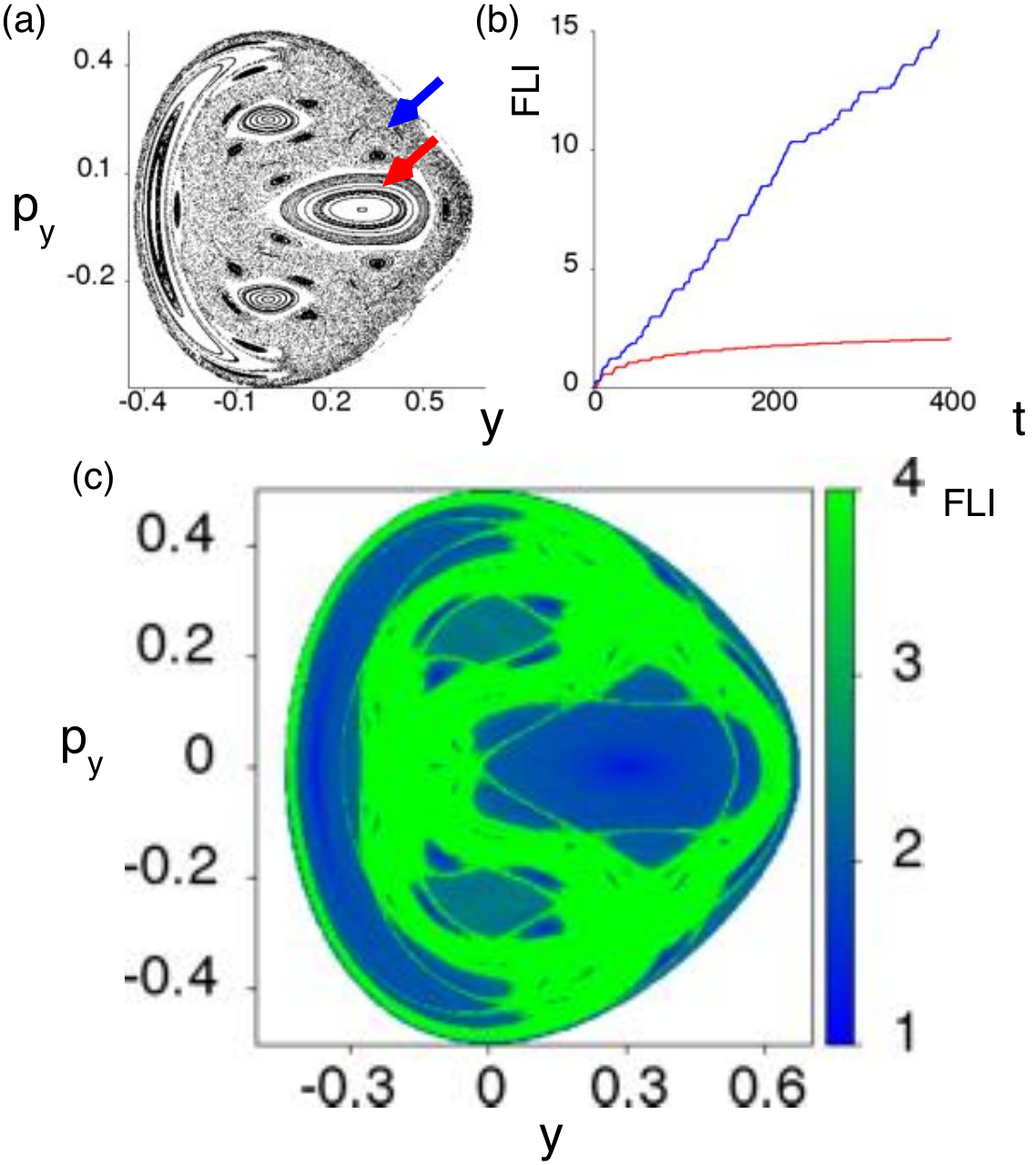}
  \caption{H\'enon-Heiles system: (a) The $(y,p_y)$ surface of section at a total energy $E = \frac{1}{8}$. The section is defined by $x = 0$ and $p_x > 0$. (b) The FLI as a function of time for the two example initial conditions indicated by arrows in (a). (c) The surface of section defined in (a) is mapped using the FLI. Total integration time $t_f = 300$. Note that values of FLI $\geq 4$   are shown in green.}
  \label{fgr:FLI_HH}
\end{figure}

Figure~\ref{fgr:FLI_HH} shows the results of the FLI computation for the H\'enon-Heiles system. The phase space at a total energy $E=\frac{1}{8}$ has a mixed regular-chaotic nature, as shown in Fig.~\ref{fgr:FLI_HH}(a) using an appropriate surface of section.  In the regular region, the classical dynamics is stable and localized for all time. In the regular region, according to the KAM theorem, one can define actions which are constants of the motion. On the other hand, the dynamics is unstable in the chaotic region, i.e. the trajectories can explore the whole constant energy surface (except the regular regions) over sufficiently long times. In addition, the chaotic trajectories have only the total energy as a constant of the motion. Figure~\ref{fgr:FLI_HH}(b) shows the FLI as a function of time for two example initial conditions indicated in Fig.~\ref{fgr:FLI_HH}(a). As advertised, the FLI for the regular trajectory saturates quickly and increases almost linearly with time for the chaotic trajectory. Note that, a clear distinction between the FLI values of the two trajectories is already apparent for $t \sim 100$. Therefore, any $t_f \geq 100$ is sufficient for this system. In Figure~\ref{fgr:FLI_HH}(c) we show the FLI values associated with the various initial conditions on the surface of section.  The initial conditions are integrated for a total time $t_f = 300$ and the FLI value at the final time is projected on to the $y-p_y$ surface of section. The close correspondence between Fig.~\ref{fgr:FLI_HH}(a) and Fig.~\ref{fgr:FLI_HH}(c) is clear. The sharp difference in the FLI values clearly identify the different dynamical regions in the phase space.


\bibliographystyle{rsc} 
\bibliography{arxiv_pccp_perspective2020} 

\end{document}